\documentclass[aps,prb,showpacs,twocolumn,showpacs]{revtex4-1}

\usepackage{graphicx}
\usepackage{amsmath}
\usepackage{amssymb}
\usepackage{wasysym}
\usepackage{color}

\definecolor{lila}{rgb}{0.5,0,1}
\newcommand{\bnen}{\begin{equation}}
\newcommand{\eden}{\end{equation}}
\newcommand{\bean}{\begin{eqnarray}}
\newcommand{\eean}{\end{eqnarray}}

\newcommand{\bna}{\begin{array}}
\newcommand{\eda}{\end{array}}

\begin{document}

\title{Competition between Hund's coupling and Kondo effect in a one-dimensional extended periodic
Anderson model}

\author{I. Hagym\'asi}
\author{J. S\'olyom}
\author{\"O. Legeza}

\affiliation{Strongly Correlated Systems "Lend\"ulet" Research Group, Institute for Solid State
Physics and Optics, MTA Wigner Research Centre for Physics, Budapest H-1525 P.O. Box 49, Hungary
}

\date{\today}

\begin{abstract}
We study the ground-state properties of an extended periodic Anderson model to understand the role
of Hund's coupling between localized and itinerant electrons using the density-matrix
renormalization group algorithm. By calculating the von Neumann entropies  we show that two phase
transitions occur and two new phases appear as the
hybridization is increased in the symmetric half-filled case due to the competition between 
Kondo-effect and Hund's coupling. In the intermediate phase, which is bounded by two critical
points, we found a dimerized ground state, while in the other spatially homogeneous phases the
ground state is Haldane-like and Kondo-singlet-like, respectively. We also determine the
entanglement spectrum and the
entanglement diagram of the system by calculating the
mutual information thereby clarifying the structure of each phase. 
\end{abstract}


\maketitle

\section{Introduction}
The colossal magnetoresistance observed in certain manganites has attracted much attention in
condensed matter physics. \cite{Jin15041994,CMR_review} Typically perovskite manganites, for
example 
La$_{2/3}$Ca$_{1/3}$MnO$_3$, exhibit this enormous enhancement of the magnetoresistance.
\cite{Jin15041994} It is now generally accepted that the colossal magnetoresistance originates from
the strong correlation between electrons, namely, the double exchange and superexchange
interactions, Jahn-Teller effect etc. play an important role, although this phenomenon is far from
being understood.\cite{CMR_review1} In these compounds due to the crystal field, the five-fold
degenerate $d$ orbitals are split into two-fold degenerate ($e_g$) and three-fold degenerate
($t_{2g}$) orbitals which are coupled to each other via  Hund's coupling. The electrons in the
$e_g$ orbitals are delocalized while the $t_{2g}$ electrons are localized.\cite{PhysRevLett.74.5108}
To understand the compounds in question, it is crucial to investigate models with more than one
orbital. Hybridization between different orbitals
is claimed to be important also in other compounds like
Ca$_{2-x}$Sr$_x$RuO$_4$.\cite{PhysRevLett.84.2666} This compound exhibits several
remarkable phenomena like orbital-selective Mott transition\cite{orbmott:1}
or heavy-fermion\cite{Fisk:exp} behavior as the chemical concentration or temperature is
varied. To account for these effects, multiorbital systems have been in the focus of intensive
research recently.\cite{orbmott:2,Koga01062005,Koga:multiorbprb,Bunemann,Medici:prb}
Our purpose here is to consider further the properties of multiorbital systems.
\par The starting point of our investigations is the periodic Anderson
model,\cite{Fulde:review,Hewson:book,Patrik:book} which is a minimal model with two kinds of
electrons:
\begin{equation}   \begin{split}
   \hspace{0.2cm}  &\mathcal{H}_{\rm PAM} = 
\sum_{\boldsymbol{k}\sigma}\varepsilon_{\boldsymbol{k}}
       \hat{c}_{\boldsymbol{k}\sigma}^{\dagger}
	   \hat{c}^{\phantom \dagger}_{\boldsymbol{k}\sigma}
+\varepsilon_f\sum_{j\sigma}\hat{f}^{ \dagger}_{j\sigma}\hat{f}^{\phantom \dagger}_{j\sigma}
\\&+\sum_{j\boldsymbol{k}\sigma}(V_{\boldsymbol{k}}^{\phantom\dagger}e^{-i\boldsymbol{kR}_{j}}\hat{f
} _ { j\sigma } ^ { \dagger }
	  \hat{c}^{\phantom \dagger}_{\boldsymbol{k}\sigma}
    +V_{\boldsymbol{k}}^{*}e^{i\boldsymbol{kR}_{j}}\hat{c}_{\boldsymbol{k}\sigma}^{\dagger}
\hat{f}^{\phantom \dagger}_{j\sigma}) \\
&+U_f\sum_{j}\hat{n}^{(f)}_{j\uparrow}\hat{n}^{(f)}_{j\downarrow}.
\end{split}
\label{eq:Hamiltonian:PAM}
\end{equation}
Here
$\hat{c}_{\boldsymbol{k}\sigma}^{\dagger}$ ($\hat{c}_{\boldsymbol{k}\sigma}^{\phantom \dagger}$)
creates
(annihilates) an electron with spin $\sigma$ in a wide band with dispersion curve
$\varepsilon_{\boldsymbol{k}}$. This band will be described in the tight-binding
approximation with nearest-neighbor overlap only. Furthermore, $\hat{f}^{ \dagger}_{j\sigma}$
($\hat{f}^{\phantom \dagger}_{j\sigma}$) creates (annihilates) localized electrons at site $j$
with spin $\sigma$ and the corresponding particle number operator is
$\hat{n}^{(f)}_{j\sigma}=\hat{f}_{j\sigma}^{\dagger}\hat{f}_{j\sigma}^{\phantom\dagger}
$. They can be mixed to the states of the wide band by the matrix element $V_{\boldsymbol{k}}$.
When the states of the wide band are written in a real-space representation, this mixing is in
general non-local, however, we neglect its $\boldsymbol{k}$-dependence in the
following, $V_{\boldsymbol{k}}=V$. The Coulomb interaction between localized electrons is
 $U_{f}$, and their on-site energy is $\varepsilon_f$.
\par Our goal is to study the interactions between the two kinds of electrons, whose general form
in momentum space can be written as:
\begin{equation}   
\mathcal{H} =  \mathcal{H}_{\rm
PAM}+\frac{1}{2V}\sum\limits_{\substack{\boldsymbol{kk}^{\prime}\boldsymbol
{ q } \\\sigma\sigma^{\prime}}} U_ {
\sigma\sigma^ { \prime } } (\boldsymbol{q})c^{\dagger}_{
\boldsymbol{k}+\boldsymbol{q}\sigma}f^{\dagger}_{\boldsymbol{k}^{\prime}-\boldsymbol{q}\sigma^{
\prime } } f^ {
\phantom\dagger}_{\boldsymbol{k}^{\prime}\sigma^{\prime}}c^{\phantom\dagger}_{
\boldsymbol{k}\sigma},
\label{eq:Hamiltonian_momentum}
\end{equation}
which is visibly non-local in real-space for a general $U_ {
\sigma\sigma^ { \prime } } (\boldsymbol{q})$ interaction. In what follows we consider only on-site
interactions between the two kinds of electrons, taking into account
the interorbital Coulomb interaction and the direct exchange between them,
 described by the rotationally invariant Kanamori Hamiltonian.\cite{Medici:selective} In addition
to that we also include an on-site repulsion for the delocalized electrons. To write the
Hamiltonian in a convenient form,
 we use the real-space representation for the creation and
annihiliation operators of the electrons in the wide band ($\hat{c}^{ \dagger}_{j\sigma}$,
$\hat{c}^{ \phantom\dagger}_{j\sigma}$), then the total Hamiltonian becomes:
\begin{equation}   \begin{split}
&   \hspace{0.2cm}  \mathcal{H} =  \mathcal{H}_{\rm
PAM}+U_c\sum_{j}\hat{n}^{(c)}_{j\uparrow}\hat{n}^{(c)}_{j\downarrow}+\sum_{j\sigma\sigma^{\prime}}
(U_{cf} -J\delta_{\sigma\sigma^{\prime}}
)\hat{n}^{(c)}_{j\sigma}
\hat{n}^{(f)}_{j\sigma^{\prime}}\\
&-J\sum_{j}\left[\left(\hat{c}^{\dagger}_{j\uparrow}\hat{c}^{\phantom
\dagger}_{j\downarrow}\hat{f}^{\dagger}_{j\downarrow}\hat{f}^{\phantom
\dagger}_{j\uparrow}+\hat{c}^{\dagger}_{j\uparrow}\hat{c}^{\dagger}_{j\downarrow}\hat{f
}^{
\phantom\dagger}_{j\uparrow}\hat{f}^{\phantom\dagger}_{j\downarrow
}\right)+  {\rm H. c.}\right],\\
\end{split}
\label{eq:Hamiltonian}
\end{equation}
where $U_c$ is the Coulomb repulsion within the wide band,
$U_{cf}$ denotes the local interaction between the two kinds of electrons and $J$ is the Hund's
coupling. It is worth noting that one can always diagonalize the bilinear part of the
Hamiltonian obtaining two orthogonal bands.  However, in that case the
intraorbital interaction becomes non-local in terms of the original operators. \cite{Oles_multiband}
\par The role of the term $U_c$ and $U_{cf}$ has been examined in several
papers.\cite{Schork:Ud,Kawakami:Ud,Hagymasi:Ud,Miyake:review,DMRG:Miyake1,Hagymasi:GW,
Kawakami:charge_order,Hagymasi:Ucf} It has been shown, that $U_c$ leads to significant enhancement
of the effective mass in the Kondo regime, while $U_{cf}$ causes critical valence fluctuations in
the mixed valence regime and can lead to charge ordering in infinite spatial dimensions. The full
Hamiltonian in Eq. (\ref{eq:Hamiltonian}) including $J$ has also been
investigated thoroughly by dynamical mean-field theory (DMFT) in infinite spatial
dimensions,\cite{Koga01062005,Koga:para,Koga:magnetic,Medici:selective} and it was
revealed how the Kondo and Mott insulating states compete with the metallic state in the
half-filled case if the system is
assumed to be paramagnetic.\cite{Koga:para} It turned out, however, that magnetic long-range order
can also be present in the model, namely two types of antiferromagnetic order emerge beside the
Kondo insulating state.\cite{Koga:magnetic} The occurence of these phases originates from the
competition between  
Hund's coupling and Kondo effect. While the former one aligns the spins of localized and
itinerant electrons ferromagnetically at a given site, the latter one tries to screen the localized
spins by forming singlets with the itinerant electrons. It has also been discussed how the
hybridization affects the orbital-selective Mott localization emerging in two-orbital
Hubbard-models.\cite{Medici:selective}
\par Since the DMFT approach completely neglects the spatial fluctuations, which is only valid in
infinite dimensions, it is necessary to investigate low-dimensional systems where quantum
fluctuations are known to be much stronger. Our main goal in this paper is to explore the
one-dimensional behavior of the Hamiltonian in Eq. (\ref{eq:Hamiltonian}). In earlier
papers it has been shown\cite{Zawa:RG,Solyom:RG,Hagymasi:Ucf} that there is no
quantum phase transition in one
dimension in the absence of the Hund's coupling in contrast to the infinite dimensional case. The
competition between
the Hund's coupling and the Kondo effect may lead to the appearance of quantum phase transitions and
unexpected phases even if true long-range order cannot be present in one dimension. 
\par We apply the density-matrix
renormalization-group
method\cite{White:DMRG1,White:DMRG2,schollwock2005,manmana2005,hallberg2006} (DMRG),
which is a powerful tool to find
the ground state and to determine the correlation functions. 
Further advantage of
the DMRG method is that we can easily determine the von Neumann
entropies\cite{legeza2003b,vidallatorre03,calabrese04,rissler2006,legeza2006,luigi2008} of single
and multisite subsystems, without the need to calculate excited states, which is in general
difficult near a critical point, and their anomalies can be used to detect quantum phase
transitions.\cite{gu:prl2004,wu:prl2004,yang:pra2005,deng:prb2006}
\par In our DMRG calculation we applied the dynamic block-state selection
algorithm\cite{DBSS:cikk1,DBSS:cikk2} in which the threshold value of the quantum information loss,
$\chi$, is set a priori. We have taken  
$\chi=3\cdot10^{-6}$. A maximum of 2000 block states is needed to achieve this 
accuracy, and the truncation error was in the order of $10^{-7}$. Such low value of $\chi$ is
necessary in order to obtain 'smooth' data sets close to critical points. 
We investigated chains up to a maximum length $L=120$ with open boundary conditions and performed
8 sweeps. 
\par The setup of the paper is as follows. In Sec. II. we define the von Neumann entropies of
various subsystems used in our
analysis, and the mutual information.\cite{rissler2006,barcza2010,Boguslawski} In Sec. III. A, B and
C
we discuss the properties of the phases occuring in the model using the mutual information and the
eigenvalue spectra of the two-site density matrices. In Sec. III. D we discuss the differences
between the phase diagram obtained in the DMFT and for the one-dimensional model. Finally,
in Sec. IV. our
conclusions are presented.

\section{Von Neumann entropies}
The von Neumann entropies of different subsystems
are known to exhibit anomalies near critical points.\cite{wu:prl2004,gu:prl2004,legeza2006} 
We
examined the one-site $s_i$, two-site $s_{ij}$ entropies and the block entropy which is the entropy
 of the subsystem containing sites from 1 to $L/2$. These quantities can be obtained from the
appropriate
reduced
density matrices.\cite{legeza2003b,vidallatorre03,legeza2006} The entropy of a single site can be
obtained as
\begin{equation}
 s_i=-{\rm Tr} \rho_i\ln\rho_i,
\end{equation}
 where $\rho_i$ is the reduced density matrix of site $i$,
which is derived from the density matrix of the total system by tracing out the configurations of
all other sites. We also define the entropies  corresponding to
the two types of 
electrons at a site ($s^{(c)}_i$, $s^{(f)}_i$) in the following way:
\begin{align}
 s_i^{(c)}=-{\rm Tr} \rho_i^{(c)}\ln\rho_i^{(c)},\\
 s_i^{(f)}=-{\rm Tr} \rho_i^{(f)}\ln\rho_i^{(f)}, 
\end{align}
where $\rho_i^{(c)}$ ($\rho_i^{(f)}$) is obtained by performing an additional trace over the
remaining $f$ ($c$) degrees of freedom at site $i$. The two-site entropy is written as
\begin{equation}
 s_{ij}=-{\rm Tr}
\rho_{ij}\ln\rho_{ij},
\end{equation}
where $\rho_{ij}$ is the two-site reduced density matrix of sites $i$ and
$j$. We can also introduce the partial two-site entropies for  type $a$ electrons on
site $i$
and type $b$  electrons on site $j$:
\begin{equation}
 s_{ij}^{(ab)}=-{\rm Tr}
\rho_{ij}^{(ab)}\ln\rho_{ij}^{(ab)}, \quad a,b\in\{c,f\}
\end{equation}
where $\rho_{ij}^{(ab)}$ is derived from $\rho_{ij}$ by tracing out the states of the other
electrons. 
The mutual information\cite{wolf2008,furukawa2009,legeza:entanglement} which measures the
entanglement between sites $i$ and $j$ can be obtained from:
\begin{gather}
   I_{ij}=s_i+s_j-s_{ij},
\end{gather}
while the mutual
information between $a$ and $b$ type electrons on sites $i$ and $j$ is defined as
\begin{gather}
   I_{ij}^{(ab)}=s_i^{(a)}+s_j^{(b)}-s_{ij}^{(ab)},
\end{gather}
which measures all correlations both of classical and quantum origin between $a$ and $b$
type electrons on sites $i$ and $j$. In what follows we refer to
$I_{ij}^{(ab)}$ as the entanglement between these components.  Finally, the block entropy
is defined as
\begin{gather}
s(L/2)=-{\rm Tr} \rho_A\ln\rho_A,
\end{gather}
where $A$ denotes the subsystem which contains the sites from 1 to $L/2$. In contrast to the one- or
 two-site entropies, which have a finite upper bound, the block entropy grows as $\mathcal{O}(\ln
L)$ for one-dimensional critical systems.\cite{calabrese04,vidallatorre03}

\section{Results}
In what follows we consider the half-filled case and nearest-neighbor hopping between
delocalized electrons, $\varepsilon(k)=-2t\cos k$, and use the half bandwidth,
$W=2t$, as the 
energy scale of the system and set $\varepsilon_f=0$. For simplicity we assume
$U_c=U_f=U$ and
$U=U_{cf}+2J$.
In the absence of Hund's coupling the ground state in one dimension is either a collective singlet
or consists of
less entangled local Kondo-singlets depending on the values of Coulomb interactions and the
hybridization.\cite{Hagymasi:Ucf} There is no quantum phase transition between these phases, just a
smooth crossover
separates them. To examine the effect of the Hund's coupling, firstly we consider what happens for a
finite Hund's coupling, namely for $J/U=0.1$ and $0.3$, with $U=4W$
as the hybridization is varied.
\par
\begin{figure}[!htb]
\includegraphics[scale=0.6]{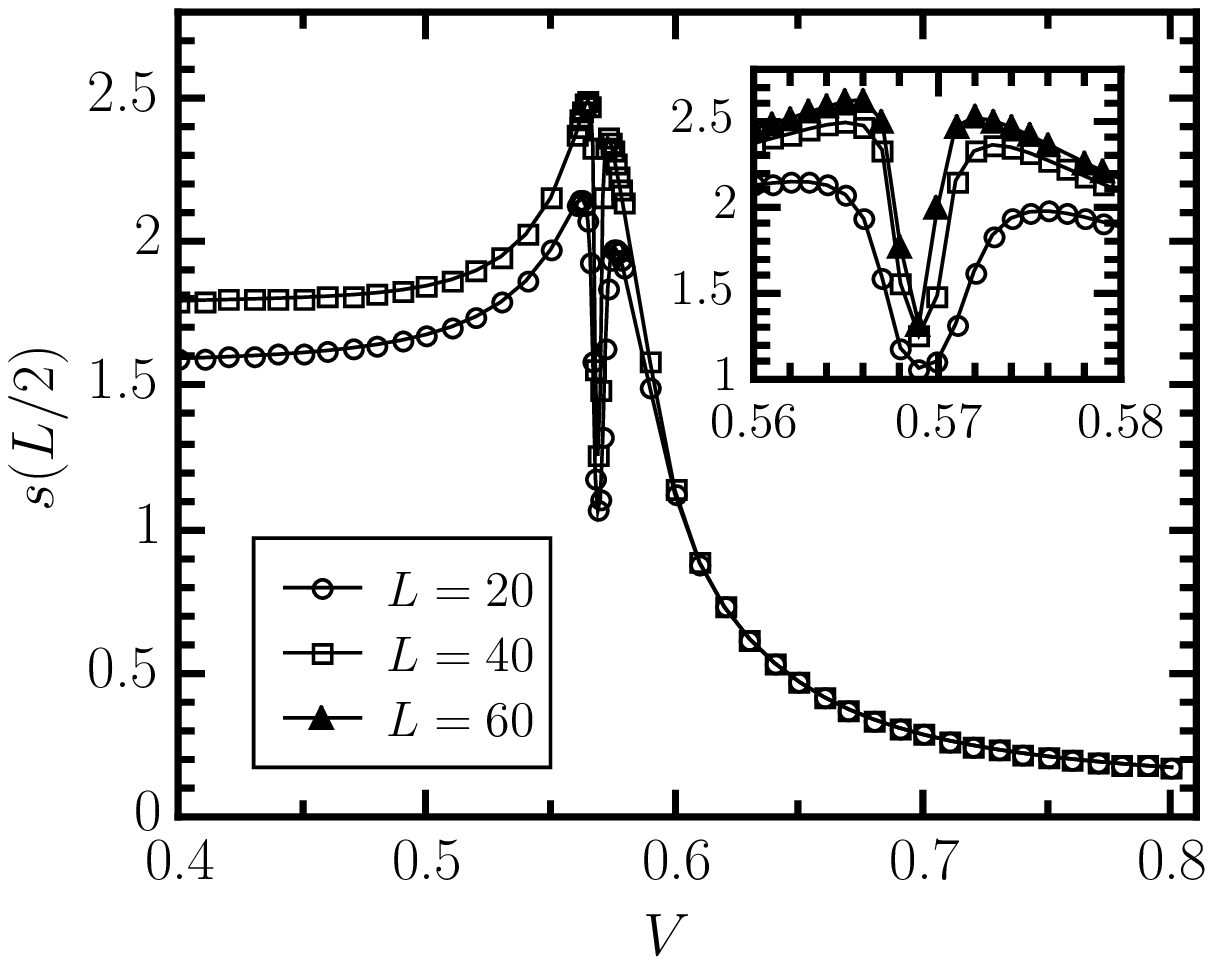}
\caption{The block entropy, $s(L/2)$ as a function of hybridization for different chain lengths and
$J/U=0.1$. The lines are guides to the eye.}
\label{fig:block_entropy}
\end{figure}

\begin{figure}[!htb]
\includegraphics[scale=0.6]{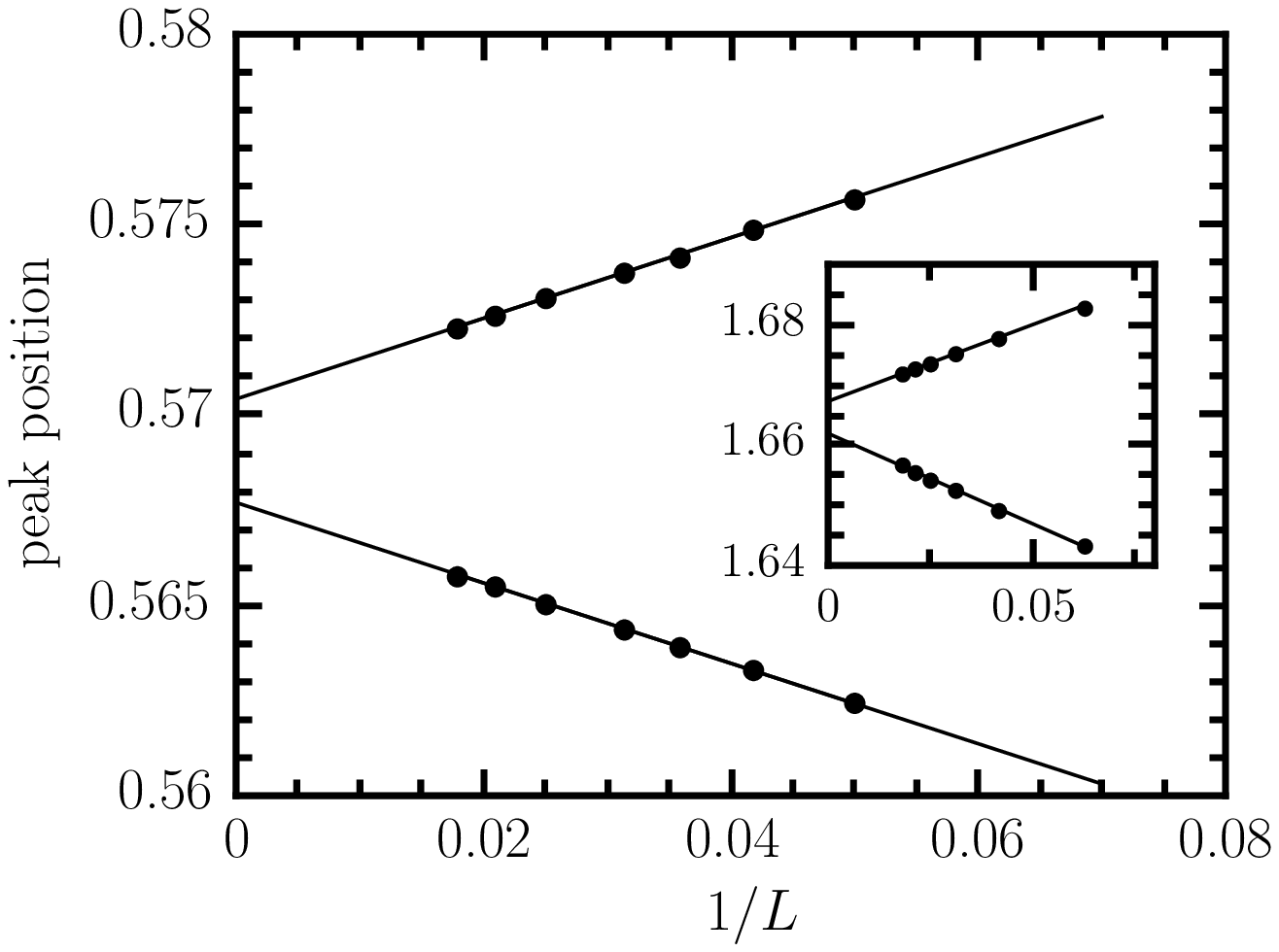}
\caption{The main figure shows the finite-size scaling of the peaks occurring in the block entropy
for $J/U=0.1$, while the inset shows the same for $J/U=0.3$. The lines are the best linear fits to
the data. The estimated error is comparable to the size of the symbols.}
\label{fig:peak_pos_scaling}
\end{figure}

 Firstly, we investigate the block entropy of one half of the chain.
This quantity is a smooth function of $V$ for any $U$ when $J=0$. For any finite $J$, however, two
peaks appear in the block entropy as can be seen in Fig. \ref{fig:block_entropy} for different chain
lengths for a fixed value of $J/U=0.1$, where the two peaks are around  $V/W=0.57$.
 It is clearly observed that
the height of the peaks increases as the chain length
is increased. We know that maxima in
the
block entropy can be attributed to quantum critical
points\cite{legeza2006} if they evolve into anomalies in the thermodynamic limit.  Two peaks may
indicate the existence of two phase transitions separating three different phases. To
check if it is indeed the case, one has to show that the peaks remain separated and do not merge in
the thermodynamic limit. The finite-size scaling of the position of the peaks is shown in Fig. 
\ref{fig:peak_pos_scaling}. We could treat systems with $L=60$ sites near the critical points due
to the high value of the block entropy. To determine the positions of the maxima accurately we used
a cubic
spline interpolation.
Figure \ref{fig:peak_pos_scaling} shows that the position of the peaks as a function of $1/L$ can be
fitted well
with a linear  function and the phases remain separated in the thermodynamic limit having positions:
\begin{equation}
\begin{split}
V^{\rm cr}_1/W=0.5677(0)\\
V^{\rm cr}_2/W=0.5704(1)
\end{split}
\end{equation} 
for $J/U=0.1$. As a remark we mention that to estimate the error of the data points in Fig.
\ref{fig:peak_pos_scaling} we use the following analysis. The quantum information loss was a priori
set to 
$\chi=3\cdot10^{-6}$. Therefore the relative error of the ground state energy is estimated to be of
the same order 
of magnitude, $\delta E_{\rm rel}\sim 10^{-6}$. Since the ground state energies are of the order of
$E\sim 10^2$, 
the absolute error of the energies is of the order of $\delta E_{\rm abs}\sim 10^{-4}$, so the error
of the block 
entropy can be expected to be not larger than $\delta S_{\rm abs}\sim 10^{-4}$. The block entropy
values near the critical 
points were sampled with an equidistant step $5\cdot10^{-4}$ (for better visibility not all are
shown in Fig. \ref{fig:block_entropy}). 
The error of the spline fit is also expected to be $~10^{-4}$. All in all the overall error is
expected to be $~10^{-4}$, 
which is comparable to the size of the symbols in Fig. \ref{fig:peak_pos_scaling}. 

We repeated the same  calculation for other values of $J/U$.  Our results indicate that  two
separate peaks are present in the block entropy for any
finite $J$. The distance of the peaks is shown in Table
\ref{table:0} for two values of $J/U$ for different chain lengths.
\begin{table}[!htb]
\begin{tabular}{@{}c@{\hspace{4mm}}c@{\hspace{4mm}}c@{\hspace{4mm}}c@{\hspace{4mm}}}
\toprule
$J/U$  & $L=16$  & $L=24$ & $L\to\infty$\tabularnewline\hline
$0.1$  & $0.016$  & $0.011$ & $0.0027(1)$\tabularnewline
$0.3$  & $0.040$  & $0.028$ & $0.006(8)$\tabularnewline\hline\hline
\end{tabular}
\caption{The distance of the peaks of the block entropy for several chain lengths and $J/U$
ratios. The extrapolation was performed using chains up to $L=60$.}
\label{table:0} 
\end{table}
It is clear that their distance grows as $J$ is increased, furthermore, the critical values of $V$
are also shifted towards larger values.
In what follows we analyze the ground state properties of each phase for $J/U=0.1$ using the mutual
information and correlation functions to check if the peaks indeed separate different phases.

\subsection{The Kondo singlet phase for $V^{\rm cr}_2<V$}
\begin{figure}[!htb]
\includegraphics[scale=0.5]{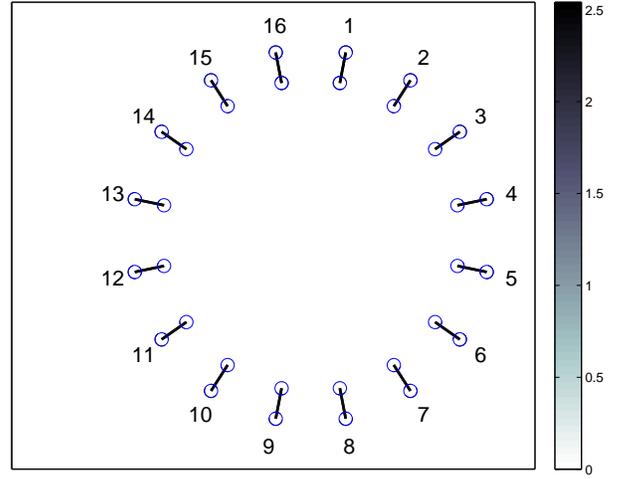}
\caption{Schematic view of all components of the mutual information ($I_{ij}^{(cc)}$,
$I_{ij}^{(cf)}$, $I_{ij}^{(ff)}$) for $V/W=0.8$ and $L=16$. The numbers are the site indices. The
inner and  outer circles denote type $f$ (localized) and type $c$ (itinerant) electrons.}
\label{fig:mutual_inf_V0p8}
\end{figure}

Firstly, we consider what happens for large hybridization where the effect of Hund's coupling is
expected to be small compared to that of hybridization and the Coulomb interaction and the
properties of the $J=0$ model are expected to be recovered. We examine how the
individual system components are entangled to each other using
the mutual  information. We have seen already
in Fig. \ref{fig:block_entropy} that for large values of the hybridization the block entropy
decreases rapidly, which indicates a less entangled state. This is the case indeed, as is seen in
the entanglement map in Fig. \ref{fig:mutual_inf_V0p8}.
We can see that very strong on-site entanglement appear, while the one-particle states on different
sites are hardly entangled to each other. To describe the physical origin of this difference in the
entanglement within a
site and between neighboring sites we calculated the eigenvalues ($\omega_{\gamma}$,
$\gamma=1,\dots,16$) of the two-site
density matrix $\rho_{ij}^{(ab)}$ and the corresponding eigenfunctions. From
$\rho_{L/2,L/2}^{(cf)}$ we found that for $V/W=0.8$ one of its eigenvalues is larger by two orders
of magnitude
than
the others and the corresponding eigenvector is:
\begin{equation}
\label{eq:wavefunction:V0p8_onsite}
\begin{split}
 \phi^{(cf)}_{L/2,L/2}&=
  0.5574(|\uparrow\rangle_{L/2}^{(c)}|\downarrow\rangle_{L/2}^{(f)}
 -|
\downarrow\rangle_{L/2}^{(c)}|\uparrow\rangle_{L/2}^{(f)})\\
& + 0.4350(|\uparrow\downarrow\rangle_{L/2}^{(c)}|0\rangle_{L/2}^{(f)}
+|0\rangle_{L/2}^{(c)}
|\uparrow\downarrow\rangle_{L/2}^{(f)}).
\end{split}
\end{equation}
Here $|0\rangle_{i}^{a}$, $|\uparrow\rangle_{i}^{a}$, $|\downarrow\rangle_{i}^{a}$,
$|\uparrow\downarrow\rangle_{i}^{a}$ denote the four possible states of electron type
$a$ on site $i$.
We can see that strong on-site singlets are formed between localized and delocalized electrons,
which we may refer to as Kondo-singlets, since this is the consequence of the enhanced Kondo effect.
 The entanglement between nearest-neighbor sites is smaller by two orders of magnitude than the
on-site entanglement between localized and delocalized electrons. Therefore, the ground state is
almost a product state. Since the eigensystem of $\rho_{L/2,L/2+1}^{(cc)}$ and
$\rho_{L/2,L/2+1}^{(ff)}$ is quantitatively the same, we consider only the former one. The
eigenfunction belonging to the most significant eigenvalue of $\rho_{L/2,L/2+1}^{(cc)}$ reads:
\begin{equation}
\label{eq:wavefunction:V0p8_nn}
\begin{split}
 &\phi^{(cc)}_{L/2,L/2+1}=\\
  &0.6900(|\uparrow\rangle_{L/2}^{(c)}|\downarrow\rangle_{L/2+1}^{(c)}
 -|
\downarrow\rangle_{L/2}^{(c)}|\uparrow\rangle_{L/2+1}^{(c)})\\
& + 0.1545(|\uparrow\downarrow\rangle_{L/2}^{(c)}|0\rangle_{L/2+1}^{(c)}
+|0\rangle_{L/2}^{(c)}
|\uparrow\downarrow\rangle_{L/2+1}^{(c)}),
\end{split}
\end{equation}
that is, the nearest neighbor coupling  between the spins is antiferromagnetic.
We checked that the mutual information components
have  their bulk values at $L=16$ already, which is due to the hardly entangled ground state.
Indeed, the properties of this phase agree with the known behavior of the conventional periodic
Anderson model for large hybridization.

\subsection{The Haldane-like phase for $V<V_{1}^{\rm cr}$}
\begin{figure}[!htb]
\includegraphics[scale=0.5]{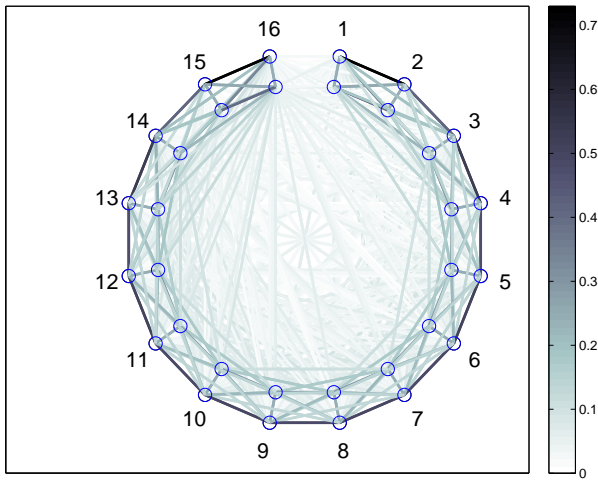}
\caption{The same as in Fig. \ref{fig:mutual_inf_V0p8}, but for $V/W=0.3$.}
\label{fig:mutual_inf_V0p3}
\end{figure}

A new phase is expected to appear for small hybridization, where $J$ dominates.
Here we discuss the properties of the phase emerging for $V<V_{1}^{\rm cr}$ using the mutual
information. The entanglement diagram containing all types of the mutual information is
shown in Fig. \ref{fig:mutual_inf_V0p3} for $V/W=0.3$. 
One can see that the strongest entanglement is developed between neighboring delocalized electrons
and
moderately  strong entanglement is present between more distant sites.
\par Firstly, we
consider how the
 eigenvalue spectrum looks like for $\rho_{L/2,L/2}^{(cf)}$ and $L=16$. We found that one of its
eigenvalues is threefold degenerate and larger by an order of magnitude than the others. Its value
is very close to $1/3$ and the three corresponding eigenfunctions, $\phi_{L/2,L/2}^{(cf),\gamma}$
$\gamma=1,2,3$, read:
\begin{equation}
\label{eq:wavefunction:V0p3_onsite}
\begin{split}
 \phi^{(cf),1}_{L/2,L/2}=& \ |\uparrow\rangle_{L/2}^{(c)}|\uparrow\rangle_{L/2}^{(f)},\\
 \phi^{(cf),2}_{L/2,L/2}=& \
\frac{1}{\sqrt{2}}(|\uparrow\rangle_{L/2}^{(c)}|\downarrow\rangle_{L/2}^{(f)}+|
\downarrow\rangle_{L/2}^{(c)}|\uparrow\rangle_{L/2}^{(f)}\rangle),\\
 \phi^{(cf),3}_{L/2,L/2}=& \ |\downarrow\rangle_{L/2}^{(c)}|\downarrow\rangle_{L/2}^{(f)}.
\end{split}
\end{equation}
That is, the electrons on the same site are in a state where the $S=1$ triplet 
components have the largest weights. 
\par
\begin{figure}[!htb]
\includegraphics[scale=0.6]{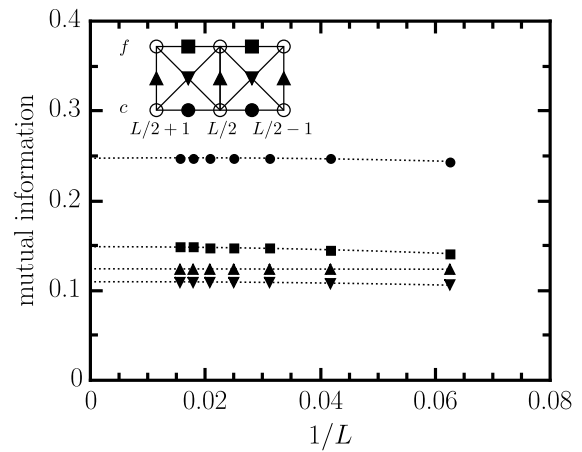}\
\caption{The finite-size scaling of the mutual information for $V=0.3W$. The dotted lines are
quadratic  polynomial fits to the data. The inset drawing denotes a segment of the middle of the
chain, the circles denote the type $c$ and type $f$ electrons on a given site, while
the symbols
denote the bonds shown in the main plot. }
\label{fig:bond_scaling_V0p3}
\end{figure}

 As a next step we examine the entanglement between nearest neighbor sites. Since the
eigensystems of
$\rho_{L/2,L/2+1}^{(cc)}$,  $\rho_{L/2,L/2+1}^{(cf)}$ and  $\rho_{L/2,L/2+1}^{(ff)}$ are
quantitatively
very similar, we only present results for $\rho_{L/2,L/2+1}^{(ff)}$. The eigenfunction corresponding
to the most significant eigenvalue is:
\begin{equation}
\label{eq:wavefunction:V0p3_nn}
\begin{split}
 &\phi^{(ff)}_{L/2,L/2+1}= \\
 &\ 0.7071(|\uparrow\rangle_{L/2}^{(f)}|\downarrow\rangle_{L/2+1}^{(f)}
 -|
\downarrow\rangle_{L/2}^{(f)}|\uparrow\rangle_{L/2+1}^{(f)})\\
& + 0.0014(|\uparrow\downarrow\rangle_{L/2}^{(f)}|0\rangle_{L/2+1}^{(f)}
+|0\rangle_{L/2}^{(f)}|
\uparrow\downarrow\rangle_{L/2+1}^{(f)}),
\end{split}
\end{equation}
which means that the entanglement between the neighboring  sites results mainly from the singlet 
formation. 
Furthermore, we investigated how
the mutual information components scale as the system size is increased. This is shown in
Fig.
\ref{fig:bond_scaling_V0p3}.
It is obviously seen that the bonds hardly change as the chain length becomes larger. The
extrapolation  was performed using a quadratic polynomial:
\begin{equation}
I_{ij}^{(ab)}(L)=I_{ij}^{(ab)}+A/L+B/L^2,
\label{eq:bond_quadratic_fit}
\end{equation}
where $I_{ij}^{(ab)}$, $A$ and $B$ are free parameters.
\par This confirms that the spins are aligned ferromagnetically
within a site, but they couple antiferromagnetically between nearest-neighbor sites. The former one
is a consequence of the strong Hund's coupling which prefers parallel alignment of the spins, while
the latter one is due to the RKKY-interaction mediated by the conduction electrons.  The same
structure was confirmed  for $V=0$. 
These findings suggest that the model in this regime can be considered as an $S=1$ Heisenberg chain
with antiferromagnetic nearest-neighbor coupling, therefore this is a Haldane-like phase. The ground
state is a singlet, however, in the thermodynamic
limit the ground state of an open chain becomes degenerate with the first $S=1$ excited state due
to the end spins. This is a well-known property of the Haldane phase.\cite{White:S1chain}

\subsection{The dimerized phase for $V^{\rm cr}_1<V<V^{\rm cr}_2$}
\begin{figure}[!htb]
\includegraphics[scale=0.5]{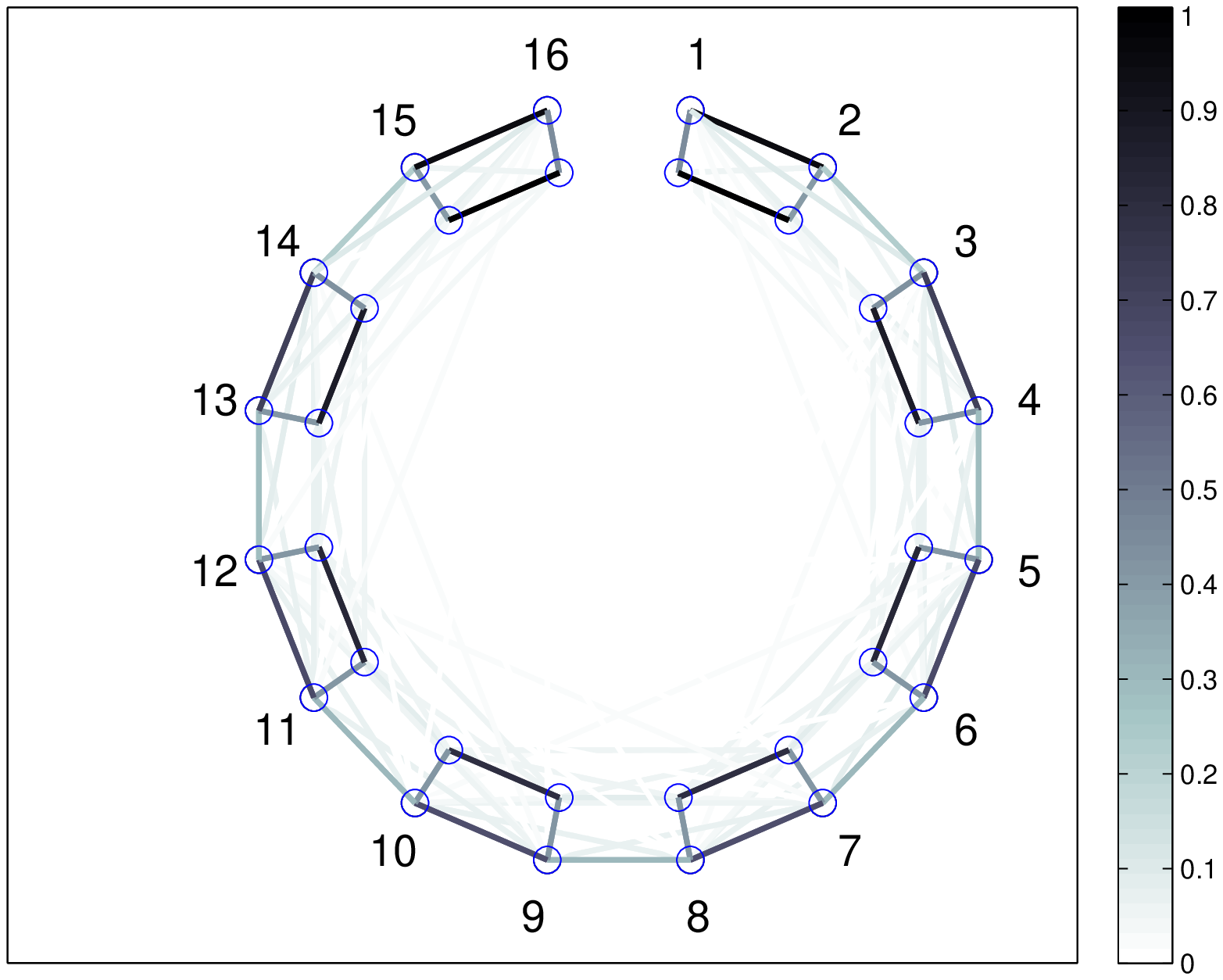}
\caption{The same as in Fig. \ref{fig:mutual_inf_V0p8}, but for $V/W=0.569$.}
\label{fig:mutual_inf_V0p569}
\end{figure}

\begin{figure}[!htb]
\includegraphics[scale=0.6]{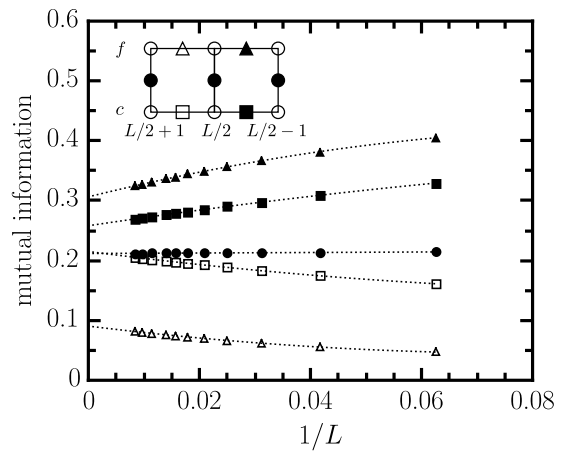}
\caption{The same as in Fig. \ref{fig:bond_scaling_V0p3}, but for $V/W=0.569$.}
\label{fig:bond_scaling_V0p569}
\end{figure}

Finally, we examine the properties of the narrow intermediate phase, whose appearance is indicated
by the analysis
of the block entropy.  Using the tools applied in the previous subsections we examine the spatial
structure of the ground state. The entanglement map shown in Fig. \ref{fig:mutual_inf_V0p569},
 is drastically different from that in the previous phase.
It is clearly seen that strong and weak bonds alternate along the chain, which suggests a spatially
inhomogeneous, dimerized ground state. Before investigating the physical processes that contribute
to the 
creation of the strong entanglement, it is important to check if the dimerization remains
finite
in the thermodynamic limit. We can introduce two types of order parameters for the dimerization:
\begin{align}
\label{eq:dim_ord_local}
D_1^{(a)}(V)&=\lim_{L\to\infty}\left|I^{(aa)}_{L/2,L/2+1}(V)-I^{(aa)}_{L/2-
1, L/2 } (V)\right| , \\
D_2(V)&=\lim_{L\to\infty}\left|s(L/2,V)-s(L/2+1,V)\right|.
\label{eq:dim_ord_nonlocal}
\end{align}
Since $D_1^{(a)}(V)$ is a local quantity, we expect that it is
less
sensitive to the boundary effects. It requires, however, the calculation of several
correlation functions, and their computation time scales as $L^2$ which can be computationally
demanding.
The quantity in Eq. (\ref{eq:dim_ord_nonlocal}) is computationally less demanding, but since the
block entropy is a non-local quantity its convergence to the bulk value may be slower. Instead of
showing $D_1^{(a)}(V)$ directly, we plot the individual values of the mutual
information components ($I^{(aa)}_{L/2,L/2+1}$, $I^{(aa)}_{L/2-1,L/2}$) and
investigate
their size-dependence. In this case we could consider chains up to $L=120$, since in the
intermediate phase the block
entropy has a much lower value than near the critical points and its low value also indicates a less
entangled ground state as expected for a dimerized phase. This is shown in Fig.
\ref{fig:bond_scaling_V0p569}.
In Fig. \ref{fig:bond_scaling_V0p569} we used a quadratic polynomial, Eq.
(\ref{eq:bond_quadratic_fit}) for the extrapolation, which gives an upper bound for the bond
strengths. 
The data can also be fitted using a power-law function:
\begin{equation}
I_{ij}^{(ab)}(L)=I_{ij}^{(ab)}+A/L^{B},
\end{equation}
where $I_{ij}^{(ab)}$, $A$ and $B$ are free parameters. The residual sum of squares is
roughly of the same order of magnitude
for both types of fits, 
 $\mathcal{O}(10^{-6})$, therefore we give the values of the order parameters for both 
fits in Table \ref{table:I}.
\begin{table}[h]
\begin{tabular}{@{}c@{\hspace{2mm}}c@{\hspace{2mm}}c@{\hspace{2mm}}c@{\hspace{2mm}}}
\toprule
type of fit &   $D_1^{(c)}(V)$  & $D_1^{(f)}(V)$ & $D_2(V)$  \\ \hline
quadratic  & $0.043(0)$& $0.215(9)$ & $0.32(4)$ \\
power-law & $0.03(1)$& $0.17(8)$ & $0.26(0)$\\
\toprule
\end{tabular}
\caption{The extrapolated dimerization order parameters for different types of fits at
$V/W=0.569$.}
\label{table:I}
\end{table}
The quadratic extrapolation clearly overestimates the order parameters while the polynomial fit
underestimates it, since we expect that the order parameter begins to saturate as soon as the bulk
limit is achieved. 
\par
\begin{figure}[!htb]
\includegraphics[scale=0.6]{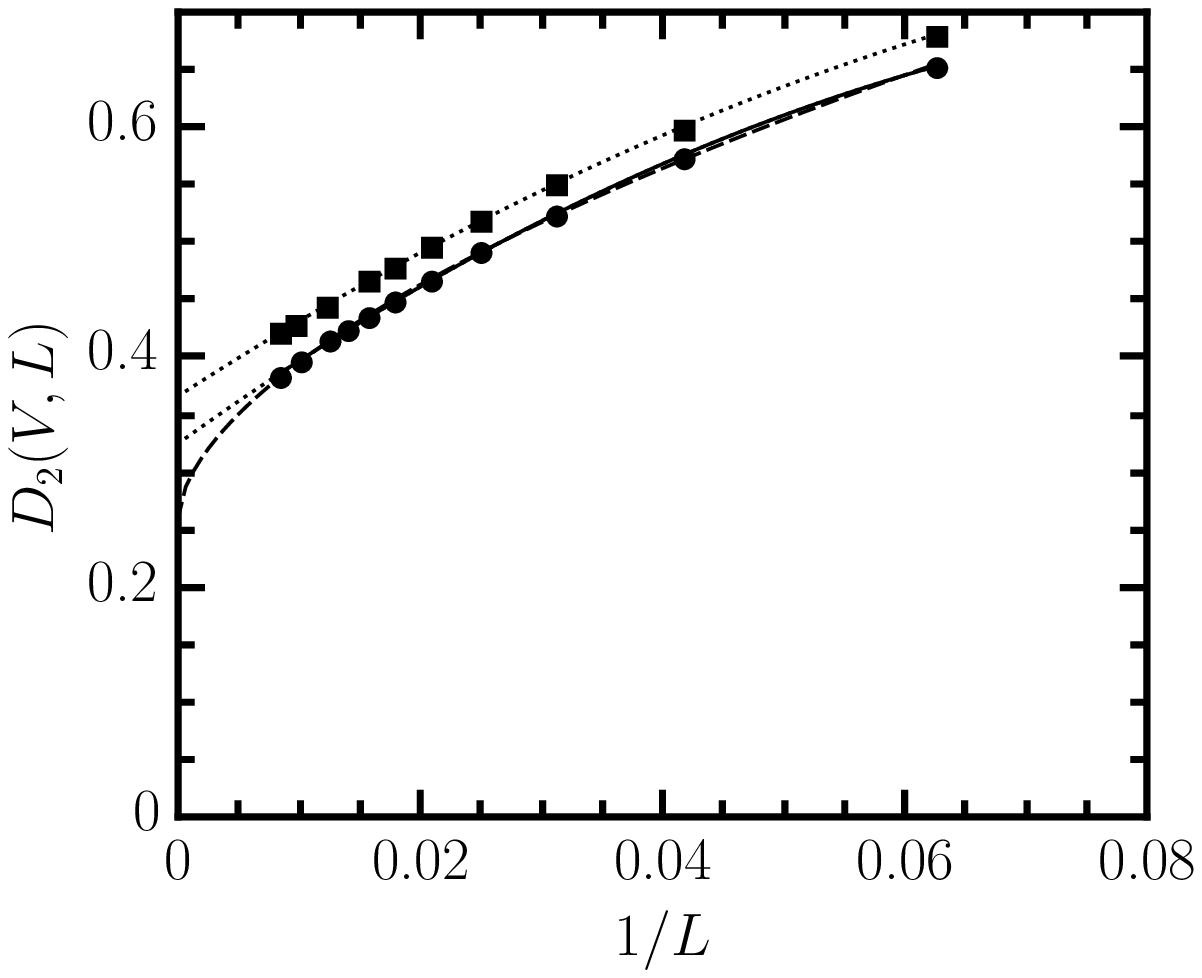}
\caption{Finite-size scaling of the order parameter in Eq. (\ref{eq:dim_ord_nonlocal}). The symbols
$\CIRCLE$, $\blacksquare$ belong to $V/W=0.569$, $J/U=0.1$ and $V/W=1.665$, $J/U=0.3$, respectively.
The dotted and dashed lines denote the quadratic and power-law fits,
respectively.}
\label{fig:entropy_scaling_V0p569}
\end{figure}

 The other order parameter defined in Eq. (\ref{eq:dim_ord_nonlocal}) is shown in Fig.
\ref{fig:entropy_scaling_V0p569} for $J/U=0.1$ and $J/U=0.3$.
 The extrapolations using the
two different fits
are shown in Table  \ref{table:I}. From these calculations we conclude that the intermediate
phase remains dimerized in the thermodynamic limit, which is a clear sign of spontaneous symmetry breaking. Moreover, it is
a surprising phenomenon, since dimerization has not been observed at all in the periodic Anderson
model without the Hund's coupling. 
\par Now, we return to  the entanglement patterns in Fig.
\ref{fig:mutual_inf_V0p569}. We consider first the entanglement between localized electrons
by examining $\rho_{L/2-1,L/2}^{(ff)}$ and $\rho_{L/2,L/2+1}^{(ff)}$ for $L=16$. We have seen that
due to
the finite-size effects the strengths of the bonds change, but the qualitative picture, which can be
obtained from the analysis of the density matrices, remains. For $\rho_{L/2-1,L/2}^{(ff)}$ one of
the
eigenvalues is larger by an order of magnitude than the others, and the corresponding eigenfunction
is:
\begin{equation}
\label{eq:wavefunction:V0p569_strong_ff}
\begin{split}
 &\phi^{(ff)}_{L/2-1,L/2}=\\
 & 0.7071(|\uparrow\rangle_{L/2-1}^{(f)}|\downarrow\rangle_{L/2}^{(f)}
 -|\downarrow\rangle_{L/2-1}^{(f)}|\uparrow\rangle_{L/2}^{(f)})\\
& + 0.006(|\uparrow\downarrow\rangle_{L/2-1}^{(f)}|0\rangle_{L/2}^{(f)}
+|0\rangle_{L/2-1}^{(f)}
|\uparrow\downarrow\rangle_{L/2}^{(f)}),
\end{split}
\end{equation}
which means that the origin of the strong entanglement between localized electrons on
neighboring sites is the
singlet formation. If we consider the neighboring bond, we obtain from $\rho_{L/2,L/2+1}^{(ff)}$
that one of the eigenvalues is $\omega_1=0.3953$ and there is a threefold degenerate eigenvalue
$\omega_2=0.1500$. The eigenvector corresponding to the former one is essentially the same as in Eq.
(\ref{eq:wavefunction:V0p569_strong_ff}), while the eigenvectors corresponding to the latter one are
the triplet components described in (\ref{eq:wavefunction:V0p3_onsite}). Due to the fact that
triplet components are mixed with a larger weight to the singlet component, it destroys the singlet
bond between the localized electrons resulting in a much weaker entanglement. Qualitatively the
above considerations remain valid for the explanation of entanglement between the itinerant
electrons. Lastly we examine the entanglement within a site with the help of
$\rho_{L/2,L/2}^{(cf)}$.
In this case we have again a non-degenerate eigenvalue, $\omega_1=0.2467$, and a threefold
degenerate one, $\omega_2=0.2274$. The eigenvector corresponding to $\omega_1$ is
\begin{equation}
\label{eq:wavefunction:V0p569_onsite}
\begin{split}
 \phi^{(cf)}_{L/2,L/2}&= \ 0.5885(|\uparrow\rangle_{L/2}^{(c)}|\downarrow\rangle_{L/2}^{(f)}
 -|\downarrow\rangle_{L/2}^{(c)}|\uparrow\rangle_{L/2}^{(f)})\\
& + 0.3921(|\uparrow\downarrow\rangle_{L/2}^{(c)}|0\rangle_{L/2}^{(f)}
+|0\rangle_{L/2}^{(c)}
|\uparrow\downarrow\rangle_{L/2}^{(f)}),
\end{split}
\end{equation}
while the eigenvectors of $\omega_2$ are the triplet states in (\ref{eq:wavefunction:V0p3_onsite}).
It can be seen easily, that the on-site
spin correlation is still ferromagnetic, but significantly reduced 
compared to the Haldane-like phase. While the on-site singlet state has negligible weight in the
Haldane-like phase, in the dimerized state the on-site triplet and singlet states are mixed with
comparable weights.

\subsection{Discussion}
In the light of the above results it is worth examining the nature of the phase transitions and
comparing the properties of the phases to what has been obtained in infinite dimensions.
\par As we mentioned, there is no phase transition when $J=0$, where the ground state is
Kondo-singlet-like discussed in Sec. III. A. For any finite $J$ two new phases appear, namely, a
Haldane-like and a dimerized phase whose properties are discussed below, and they disappear as
$J\to0$. 
\par We have
seen that for $V<V^{\rm cr}_1$ the ground state is Haldane-like while for $V>V^{\rm cr}_2$
Kondo-singlet-like, and both are gapful and homogeneous. For $V^{\rm
cr}_1<V<V^{\rm cr}_2$ the translational symmetry is broken due to the dimerization. 
It is worth noting that similar phase diagram has been obtained in frustrated spin
ladders,\cite{legeza:ladder} where on-site and nearest-neighbor antiferromagnetic couplings compete
with each other. 
\par
\begin{figure}[!htb]
\includegraphics[width=0.9\columnwidth]{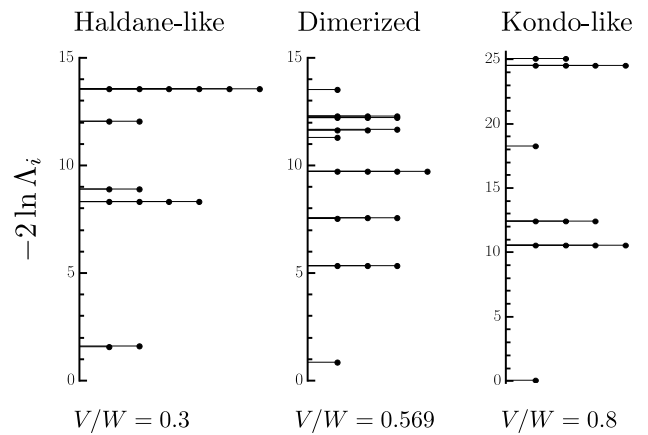}
\caption{The entanglement spectrum in the three different phases for $J/U=0.1$ and $L=120$ (for
better visibility only the lower part is shown). }
\label{fig:entanglement_spect}
\end{figure}

 The existence of different phases can be corroborated by investigating the entanglement
spectrum\cite{Haldane:entanglement} which is known to be a useful tool to detect symmetry-protected
topological order.\cite{Pollmann:entanglement} It is obtained from the eigenvalues
($\Lambda_i$) of the density matrix of a half chain. In the Haldane-like phase the degeneracy of
each eigenvalue is an even number,\cite{Pollmann:entanglement} while both even and odd degeneracies
may appear in other phases. For short chains both even and odd degeneracies occur for any $V/W$.
This picture drastically changes for
longer chains, namely, for $L>90$. Here the correlation between the end spins suddenly vanishes,
which is
marked by a jump in the block entropy for small $V/W$. While even and odd degeneracies remain in the
spectrum for $V>V_1^{\rm cr}$, 
the entanglement spectrum contains nearly degenerate eigenvalues
whose multiplicity is even for $V<V_1^{\rm cr}$.  In the thermodynamic limit the eigenvalues becomes
exactly degenerate.
In Fig. \ref{fig:entanglement_spect} the low-lying eigenvalues and their degeneracies are shown with
a logarithmic scale corresponding to the three different phases for $L=120$. 
While the spectra of the dimerized and
Kondo-like phases contain eigenvalues with odd and even degeneracy, only even degeneracy is present
in
the Haldane-like phase in agreement with our expectation. 
\par It is interesting to compare these findings with what has been obtained by
DMFT.\cite{Koga:magnetic} Surprisingly, three distinct phases appear also in the DMFT phase diagram,
although their properties are significantly different. In DMFT there is a phase with
antiferromagnetic long-range order in which the
on-site spins are coupled ferromagnetically while the nearest-neighbor coupling is
antiferromagnetic (AF I phase). Further increase of the hybridization drives the system into
an
intermediate phase ($V^{\rm cr}_1<V<V^{\rm cr}_2$) where another type of antiferromagnetic order
takes
place. Here the on-site coupling becomes antiferromagnetic, while the nearest-neighbor
coupling remains antiferromagnetic (AF II phase). Finally, for $V^{\rm cr}_2<V$ Kondo-like behavior
is realized.
\par In one dimension we cannot expect true long-range magnetic order, only slow decay of the
correlation functions. For $V<V^{\rm cr}_1$ the on-site spins are parallel due to
the strong Hund's coupling. This Haldane-like phase might be the residue of the AF I ordered phase
obtained in the DMFT calculation. Above the second critical point,  $V^{\rm
cr}_2<V$, the ground state is homogeneous, and strong on-site correlations appear, which
originate from the enhanced Kondo-effect and the sites are occupied more and more by two localized
or delocalized electrons or vice versa. The properties of the Kondo phase are consistent with what
has
been obtained in DMFT. Both methods exhibit an intermediate phase between them, however, their
properties are completely different, which is caused by the enhanced quantum fluctuations. 
The appearance of the dimerized phase may originate from the competition between the
Haldane-like and Kondo singlet phase. This may not surprise us if we recall that dimerization has been 
found in a two-component system via a purely electronic mechanism.\cite{Sirker:dimerization} The present case
is similar although for antiferromagnetic interactions between localized electrons.

\section{Conclusions}
We investigated an extended periodic Anderson model in one dimension to understand the role of the
Hund's coupling between 
itinerant and localized electrons. We carried out accurate DMRG calculations with quantum
information loss
$\chi=3\cdot10^{-6}$ up to the accessible chain length 
$L=120$. For such error margin we found that the two competing processes, the Hund's
coupling and the Kondo effect, lead to 
the appearance of two new nontrivial phases. The enhanced quantum
fluctuations in one dimension crucially affects the properties of the phases obtained in infinite
spatial dimensions.
We performed a quantum information analysis and determined the entanglement spectrum in the various
phases and the
entanglement patterns between the
system components. Moreover, using the eigensystem of the 
two-site density matrices we examined what physical processes lead to the development of
entanglement.
We found that for  $V<V^{\rm cr}_1$ the itinerant and localized electrons form
a local triplet within a site and couple to
the nearest-neighbor sites antiferromagnetically. Here the model can be considered as an $S=1$ spin
chain with antiferromagnetic coupling and we have a Haldane-like ground state. This is also
corroborated by the entanglement spectrum. For $V^{\rm
cr}_1<V<V^{\rm cr}_2$
we have an intermediate dimerized phase, in which strong and weak singlet bonds alternate between
nearest neighbor itinerant and localized electrons. We have seen that the region where dimerization
occurs expands as the Hund's coupling is increased and shifts to large values of the hybridization.
Finally, for $V>V^{\rm
cr}_2$ the ground state is homogeneous, and local singlets are formed at each site either by  two
localized/conduction electrons or by a localized and an itinerant electron. 

\acknowledgments{This work was supported in part by the
Hungarian Research Fund (OTKA) through Grant Nos.~K 100908 and NN110360. }

\bibliography{hund_refs.bib} 

\begin{thebibliography}{10}%
\makeatletter
\providecommand \@ifxundefined [1]{%
 \ifx #1\undefined \expandafter \@firstoftwo
 \else \expandafter \@secondoftwo
\fi
}%
\providecommand \@ifnum [1]{%
 \ifnum #1\expandafter \@firstoftwo
 \else \expandafter \@secondoftwo
\fi
}%
\providecommand \enquote [1]{``#1''}%
\providecommand \bibnamefont  [1]{#1}%
\providecommand \bibfnamefont [1]{#1}%
\providecommand \citenamefont [1]{#1}%
\providecommand\href[0]{\@sanitize\@href}%
\providecommand\@href[1]{\endgroup\@@startlink{#1}\endgroup\@@href}%
\providecommand\@@href[1]{#1\@@endlink}%
\providecommand \@sanitize [0]{\begingroup\catcode`\&12\catcode`\#12\relax}%
\@ifxundefined \pdfoutput {\@firstoftwo}{%
 \@ifnum{\z@=\pdfoutput}{\@firstoftwo}{\@secondoftwo}%
}{%
 \providecommand\@@startlink[1]{\leavevmode\special{html:<a href="#1">}}%
 \providecommand\@@endlink[0]{\special{html:</a>}}%
}{%
 \providecommand\@@startlink[1]{%
  \leavevmode
  \pdfstartlink
   attr{/Border[0 0 1 ]/H/I/C[0 1 1]}%
   user{/Subtype/Link/A<</Type/Action/S/URI/URI(#1)>>}%
  \relax
 }%
 \providecommand\@@endlink[0]{\pdfendlink}%
}%
\providecommand \url  [0]{\begingroup\@sanitize \@url }%
\providecommand \@url [1]{\endgroup\@href {#1}{\urlprefix}}%
\providecommand \urlprefix [0]{URL }%
\providecommand \Eprint[0]{\href }%
\@ifxundefined \urlstyle {%
  \providecommand \doi [1]{doi:\discretionary{}{}{}#1}%
}{%
  \providecommand \doi [0]{doi:\discretionary{}{}{}\begingroup
  \urlstyle{rm}\Url }%
}%
\providecommand \doibase [0]{http://dx.doi.org/}%
\providecommand \Doi[1]{\href{\doibase#1}}%
\providecommand \bibAnnote [3]{%
  \BibitemShut{#1}%
  \begin{quotation}\noindent
    \textsc{Key:}\ #2\\\textsc{Annotation:}\ #3%
  \end{quotation}%
}%
\providecommand \bibAnnoteFile [2]{%
  \IfFileExists{#2}{\bibAnnote {#1} {#2} {\input{#2}}}{}%
}%
\providecommand \typeout [0]{\immediate \write \m@ne }%
\providecommand \selectlanguage [0]{\@gobble}%
\providecommand \bibinfo [0]{\@secondoftwo}%
\providecommand \bibfield [0]{\@secondoftwo}%
\providecommand \translation [1]{[#1]}%
\providecommand \BibitemOpen[0]{}%
\providecommand \bibitemStop [0]{}%
\providecommand \bibitemNoStop [0]{.\EOS\space}%
\providecommand \EOS [0]{\spacefactor3000\relax}%
\providecommand \BibitemShut [1]{\csname bibitem#1\endcsname}%
\bibitem{Jin15041994}%
  \BibitemOpen
  \bibfield{author}{%
  \bibinfo {author} {\bibfnamefont{S.}~\bibnamefont{Jin}}, \bibinfo {author}
  {\bibfnamefont{T.~H.}\ \bibnamefont{Tiefel}}, \bibinfo {author}
  {\bibfnamefont{M.}~\bibnamefont{McCormack}}, \bibinfo {author}
  {\bibfnamefont{R.~A.}\ \bibnamefont{Fastnacht}}, \bibinfo {author}
  {\bibfnamefont{R.}~\bibnamefont{Ramesh}},\ and\ \bibinfo {author}
  {\bibfnamefont{L.~H.}\ \bibnamefont{Chen}},\ }%
  \bibfield{journal}{%
  \bibinfo {journal} {Science}\ }%
  \textbf{\bibinfo {volume} {264}},\ \bibinfo {pages} {413} (\bibinfo {year}
  {1994})%
  \bibAnnoteFile{NoStop}{Jin15041994}%
\bibitem{CMR_review}%
  \BibitemOpen
  \bibfield{author}{%
  \bibinfo {author} {\bibfnamefont{A.~P.}\ \bibnamefont{Ramirez}},\ }%
  \bibfield{journal}{%
  \bibinfo {journal} {J. Phys.: Condens. Matter}\ }%
  \textbf{\bibinfo {volume} {9}},\ \bibinfo {pages} {8171} (\bibinfo {year}
  {1997})%
  \bibAnnoteFile{NoStop}{CMR_review}%
\bibitem{CMR_review1}%
  \BibitemOpen
  \bibfield{author}{%
  \bibinfo {author} {\bibfnamefont{Y.-K.}\ \bibnamefont{Liu}}, \bibinfo
  {author} {\bibfnamefont{Y.-W.}\ \bibnamefont{Yin}},\ and\ \bibinfo {author}
  {\bibfnamefont{X.-G.}\ \bibnamefont{Li}},\ }%
  \bibfield{journal}{%
  \bibinfo {journal} {Chin. Phys. B}\ }%
  \textbf{\bibinfo {volume} {22}},\ \bibinfo {pages} {087502} (\bibinfo {year}
  {2013})%
  \bibAnnoteFile{NoStop}{CMR_review1}%
\bibitem{PhysRevLett.74.5108}%
  \BibitemOpen
  \bibfield{author}{%
  \bibinfo {author} {\bibfnamefont{Y.}~\bibnamefont{Tomioka}}, \bibinfo
  {author} {\bibfnamefont{A.}~\bibnamefont{Asamitsu}}, \bibinfo {author}
  {\bibfnamefont{Y.}~\bibnamefont{Moritomo}}, \bibinfo {author}
  {\bibfnamefont{H.}~\bibnamefont{Kuwahara}},\ and\ \bibinfo {author}
  {\bibfnamefont{Y.}~\bibnamefont{Tokura}},\ }%
  \bibfield{journal}{%
  \bibinfo {journal} {Phys. Rev. Lett.}\ }%
  \textbf{\bibinfo {volume} {74}},\ \bibinfo {pages} {5108} (\bibinfo {year}
  {1995})%
  \bibAnnoteFile{NoStop}{PhysRevLett.74.5108}%
\bibitem{PhysRevLett.84.2666}%
  \BibitemOpen
  \bibfield{author}{%
  \bibinfo {author} {\bibfnamefont{S.}~\bibnamefont{Nakatsuji}}\ and\ \bibinfo
  {author} {\bibfnamefont{Y.}~\bibnamefont{Maeno}},\ }%
  \bibfield{journal}{%
  \bibinfo {journal} {Phys. Rev. Lett.}\ }%
  \textbf{\bibinfo {volume} {84}},\ \bibinfo {pages} {2666} (\bibinfo {year}
  {2000})%
  \bibAnnoteFile{NoStop}{PhysRevLett.84.2666}%
\bibitem{orbmott:1}%
  \BibitemOpen
  \bibfield{author}{%
  \bibinfo {author} {\bibfnamefont{V.}~\bibnamefont{Anisimov}}, \bibinfo
  {author} {\bibfnamefont{I.}~\bibnamefont{Nekrasov}}, \bibinfo {author}
  {\bibfnamefont{D.}~\bibnamefont{Kondakov}}, \bibinfo {author}
  {\bibfnamefont{T.~M.}\ \bibnamefont{Rice}},\ and\ \bibinfo {author}
  {\bibfnamefont{M.}~\bibnamefont{Sigrist}},\ }%
  \bibfield{journal}{%
  \bibinfo {journal} {Eur. Phys. J. B}\ }%
  \textbf{\bibinfo {volume} {25}},\ \bibinfo {pages} {191} (\bibinfo {year}
  {2002})%
  \bibAnnoteFile{NoStop}{orbmott:1}%
\bibitem{Fisk:exp}%
  \BibitemOpen
  \bibfield{author}{%
  \bibinfo {author} {\bibfnamefont{S.}~\bibnamefont{Nakatsuji}}, \bibinfo
  {author} {\bibfnamefont{D.}~\bibnamefont{Hall}}, \bibinfo {author}
  {\bibfnamefont{L.}~\bibnamefont{Balicas}}, \bibinfo {author}
  {\bibfnamefont{Z.}~\bibnamefont{Fisk}}, \bibinfo {author}
  {\bibfnamefont{K.}~\bibnamefont{Sugahara}}, \bibinfo {author}
  {\bibfnamefont{M.}~\bibnamefont{Yoshioka}},\ and\ \bibinfo {author}
  {\bibfnamefont{Y.}~\bibnamefont{Maeno}},\ }%
  \bibfield{journal}{%
  \bibinfo {journal} {Phys. Rev. Lett.}\ }%
  \textbf{\bibinfo {volume} {90}},\ \bibinfo {pages} {137202} (\bibinfo {year}
  {2003})%
  \bibAnnoteFile{NoStop}{Fisk:exp}%
\bibitem{orbmott:2}%
  \BibitemOpen
  \bibfield{author}{%
  \bibinfo {author} {\bibfnamefont{A.}~\bibnamefont{Koga}}, \bibinfo {author}
  {\bibfnamefont{N.}~\bibnamefont{Kawakami}}, \bibinfo {author}
  {\bibfnamefont{T.~M.}\ \bibnamefont{Rice}},\ and\ \bibinfo {author}
  {\bibfnamefont{M.}~\bibnamefont{Sigrist}},\ }%
  \bibfield{journal}{%
  \bibinfo {journal} {Phys. Rev. Lett.}\ }%
  \textbf{\bibinfo {volume} {92}},\ \bibinfo {pages} {216402} (\bibinfo {year}
  {2004})%
  \bibAnnoteFile{NoStop}{orbmott:2}%
\bibitem{Koga01062005}%
  \BibitemOpen
  \bibfield{author}{%
  \bibinfo {author} {\bibfnamefont{A.}~\bibnamefont{Koga}}, \bibinfo {author}
  {\bibfnamefont{K.}~\bibnamefont{Inaba}},\ and\ \bibinfo {author}
  {\bibfnamefont{N.}~\bibnamefont{Kawakami}},\ }%
  \bibfield{journal}{%
  \bibinfo {journal} {Prog. Theor. Phys. Suppl.}\ }%
  \textbf{\bibinfo {volume} {160}},\ \bibinfo {pages} {253} (\bibinfo {year}
  {2005})%
  \bibAnnoteFile{NoStop}{Koga01062005}%
\bibitem{Koga:multiorbprb}%
  \BibitemOpen
  \bibfield{author}{%
  \bibinfo {author} {\bibfnamefont{A.}~\bibnamefont{Koga}}, \bibinfo {author}
  {\bibfnamefont{N.}~\bibnamefont{Kawakami}}, \bibinfo {author}
  {\bibfnamefont{T.~M.}\ \bibnamefont{Rice}},\ and\ \bibinfo {author}
  {\bibfnamefont{M.}~\bibnamefont{Sigrist}},\ }%
  \bibfield{journal}{%
  \bibinfo {journal} {Phys. Rev. B}\ }%
  \textbf{\bibinfo {volume} {72}},\ \bibinfo {pages} {045128} (\bibinfo {year}
  {2005})%
  \bibAnnoteFile{NoStop}{Koga:multiorbprb}%
\bibitem{Bunemann}%
  \BibitemOpen
  \bibfield{author}{%
  \bibinfo {author} {\bibfnamefont{J.}~\bibnamefont{B{\"u}nemann}}, \bibinfo
  {author} {\bibfnamefont{D.}~\bibnamefont{Rasch}},\ and\ \bibinfo {author}
  {\bibfnamefont{F.}~\bibnamefont{Gebhard}},\ }%
  \bibfield{journal}{%
  \bibinfo {journal} {J. Phys.: Condens. Matter}\ }%
  \textbf{\bibinfo {volume} {19}},\ \bibinfo {pages} {436206} (\bibinfo {year}
  {2007})%
  \bibAnnoteFile{NoStop}{Bunemann}%
\bibitem{Medici:prb}%
  \BibitemOpen
  \bibfield{author}{%
  \bibinfo {author} {\bibfnamefont{L.}~\bibnamefont{de'Medici}}, \bibinfo
  {author} {\bibfnamefont{A.}~\bibnamefont{Georges}},\ and\ \bibinfo {author}
  {\bibfnamefont{S.}~\bibnamefont{Biermann}},\ }%
  \bibfield{journal}{%
  \bibinfo {journal} {Phys. Rev. B}\ }%
  \textbf{\bibinfo {volume} {72}},\ \bibinfo {pages} {205124} (\bibinfo {year}
  {2005})%
  \bibAnnoteFile{NoStop}{Medici:prb}%
\bibitem{Fulde:review}%
  \BibitemOpen
  \bibfield{author}{%
  \bibinfo {author} {\bibfnamefont{P.}~\bibnamefont{Fulde}}, \bibinfo {author}
  {\bibfnamefont{J.}~\bibnamefont{Keller}},\ and\ \bibinfo {author}
  {\bibfnamefont{G.}~\bibnamefont{Zwicknagl}},\ }%
  in\ \emph{\bibinfo {booktitle} {Solid State Physics: Advances in Research and
  Applications}},\ Vol.~\bibinfo {volume} {41},\ \bibinfo {editor} {edited by\
  \bibinfo {editor} {\bibfnamefont{H.}~\bibnamefont{Ehrenreich}}\ and\ \bibinfo
  {editor} {\bibfnamefont{D.}~\bibnamefont{Turnbell}}}\ (\bibinfo {publisher}
  {Academic Press},\ \bibinfo {address} {San Diego},\ \bibinfo {year} {1988})\
  pp.\ \bibinfo {pages} {1--150}%
  \bibAnnoteFile{NoStop}{Fulde:review}%
\bibitem{Hewson:book}%
  \BibitemOpen
  \bibfield{author}{%
  \bibinfo {author} {\bibfnamefont{A.~C.}\ \bibnamefont{Hewson}},\ }%
  \emph{\bibinfo {title} {The Kondo Problem to Heavy Fermions}}\ (\bibinfo
  {publisher} {Cambridge University Press},\ \bibinfo {address} {Cambridge},\
  \bibinfo {year} {1993})%
  \bibAnnoteFile{NoStop}{Hewson:book}%
\bibitem{Patrik:book}%
  \BibitemOpen
  \bibfield{author}{%
  \bibinfo {author} {\bibfnamefont{P.}~\bibnamefont{Fazekas}},\ }%
  \emph{\bibinfo {title} {Lecture notes on electron correlation and
  magnetism}}\ (\bibinfo {publisher} {World Scientific},\ \bibinfo {address}
  {Singapore},\ \bibinfo {year} {1999})%
  \bibAnnoteFile{NoStop}{Patrik:book}%
\bibitem{Medici:selective}%
  \BibitemOpen
  \bibfield{author}{%
  \bibinfo {author} {\bibfnamefont{E.~A.}\ \bibnamefont{Winograd}}\ and\
  \bibinfo {author} {\bibfnamefont{L.}~\bibnamefont{de' Medici}},\ }%
  \bibfield{journal}{%
  \bibinfo {journal} {Phys. Rev. B}\ }%
  \textbf{\bibinfo {volume} {89}},\ \bibinfo {pages} {085127} (\bibinfo {year}
  {2014})%
  \bibAnnoteFile{NoStop}{Medici:selective}%
\bibitem{Oles_multiband}%
  \BibitemOpen
  \bibfield{author}{%
  \bibinfo {author} {\bibfnamefont{A.~M.}\
  \bibnamefont{Ole\ifmmode~\acute{s}\else \'{s}\fi{}}},\ }%
  \bibfield{journal}{%
  \bibinfo {journal} {Phys. Rev. B}\ }%
  \textbf{\bibinfo {volume} {28}},\ \bibinfo {pages} {327} (\bibinfo {year}
  {1983})%
  \bibAnnoteFile{NoStop}{Oles_multiband}%
\bibitem{Schork:Ud}%
  \BibitemOpen
  \bibfield{author}{%
  \bibinfo {author} {\bibfnamefont{T.}~\bibnamefont{Schork}}\ and\ \bibinfo
  {author} {\bibfnamefont{S.}~\bibnamefont{Blawid}},\ }%
  \bibfield{journal}{%
  \bibinfo {journal} {Phys. Rev. B}\ }%
  \textbf{\bibinfo {volume} {56}},\ \bibinfo {pages} {6559} (\bibinfo {year}
  {1997})%
  \bibAnnoteFile{NoStop}{Schork:Ud}%
\bibitem{Kawakami:Ud}%
  \BibitemOpen
  \bibfield{author}{%
  \bibinfo {author} {\bibfnamefont{T.}~\bibnamefont{Yoshida}}, \bibinfo
  {author} {\bibfnamefont{T.}~\bibnamefont{Ohashi}},\ and\ \bibinfo {author}
  {\bibfnamefont{N.}~\bibnamefont{Kawakami}},\ }%
  \bibfield{journal}{%
  \bibinfo {journal} {J. Phys. Soc. Jpn.}\ }%
  \textbf{\bibinfo {volume} {80}},\ \bibinfo {pages} {064710} (\bibinfo {year}
  {2011})%
  \bibAnnoteFile{NoStop}{Kawakami:Ud}%
\bibitem{Hagymasi:Ud}%
  \BibitemOpen
  \bibfield{author}{%
  \bibinfo {author} {\bibfnamefont{I.}~\bibnamefont{Hagym\'asi}}, \bibinfo
  {author} {\bibfnamefont{K.}~\bibnamefont{Itai}},\ and\ \bibinfo {author}
  {\bibfnamefont{J.}~\bibnamefont{S\'olyom}},\ }%
  \bibfield{journal}{%
  \bibinfo {journal} {Phys. Rev. B}\ }%
  \textbf{\bibinfo {volume} {85}},\ \bibinfo {pages} {235116} (\bibinfo {year}
  {2012})%
  \bibAnnoteFile{NoStop}{Hagymasi:Ud}%
\bibitem{Miyake:review}%
  \BibitemOpen
  \bibfield{author}{%
  \bibinfo {author} {\bibfnamefont{K.}~\bibnamefont{Miyake}},\ }%
  \bibfield{journal}{%
  \bibinfo {journal} {J. Phys.: Condens. Matter}\ }%
  \textbf{\bibinfo {volume} {19}},\ \bibinfo {pages} {125201} (\bibinfo {year}
  {2007})%
  \bibAnnoteFile{NoStop}{Miyake:review}%
\bibitem{DMRG:Miyake1}%
  \BibitemOpen
  \bibfield{author}{%
  \bibinfo {author} {\bibfnamefont{S.}~\bibnamefont{Watanabe}}, \bibinfo
  {author} {\bibfnamefont{M.}~\bibnamefont{Imada}},\ and\ \bibinfo {author}
  {\bibfnamefont{K.}~\bibnamefont{Miyake}},\ }%
  \bibfield{journal}{%
  \bibinfo {journal} {J. Phys. Soc. Jpn.}\ }%
  \textbf{\bibinfo {volume} {75}},\ \bibinfo {pages} {043710} (\bibinfo {year}
  {2006})%
  \bibAnnoteFile{NoStop}{DMRG:Miyake1}%
\bibitem{Hagymasi:GW}%
  \BibitemOpen
  \bibfield{author}{%
  \bibinfo {author} {\bibfnamefont{I.}~\bibnamefont{Hagym\'asi}}, \bibinfo
  {author} {\bibfnamefont{K.}~\bibnamefont{Itai}},\ and\ \bibinfo {author}
  {\bibfnamefont{J.}~\bibnamefont{S\'olyom}},\ }%
  \bibfield{journal}{%
  \bibinfo {journal} {Phys. Rev. B}\ }%
  \textbf{\bibinfo {volume} {87}},\ \bibinfo {pages} {125146} (\bibinfo {year}
  {2013})%
  \bibAnnoteFile{NoStop}{Hagymasi:GW}%
\bibitem{Kawakami:charge_order}%
  \BibitemOpen
  \bibfield{author}{%
  \bibinfo {author} {\bibfnamefont{T.}~\bibnamefont{Yoshida}}\ and\ \bibinfo
  {author} {\bibfnamefont{N.}~\bibnamefont{Kawakami}},\ }%
  \bibfield{journal}{%
  \bibinfo {journal} {Phys. Rev. B}\ }%
  \textbf{\bibinfo {volume} {85}},\ \bibinfo {pages} {235148} (\bibinfo {year}
  {2012})%
  \bibAnnoteFile{NoStop}{Kawakami:charge_order}%
\bibitem{Hagymasi:Ucf}%
  \BibitemOpen
  \bibfield{author}{%
  \bibinfo {author} {\bibfnamefont{I.}~\bibnamefont{Hagym\'asi}}, \bibinfo
  {author} {\bibfnamefont{J.}~\bibnamefont{S\'olyom}},\ and\ \bibinfo {author}
  {\bibfnamefont{{\"O}.}~\bibnamefont{Legeza}},\ }%
  \bibfield{journal}{%
  \bibinfo {journal} {Phys. Rev. B}\ }%
  \textbf{\bibinfo {volume} {90}},\ \bibinfo {pages} {125137} (\bibinfo {year}
  {2014})%
  \bibAnnoteFile{NoStop}{Hagymasi:Ucf}%
\bibitem{Koga:para}%
  \BibitemOpen
  \bibfield{author}{%
  \bibinfo {author} {\bibfnamefont{A.}~\bibnamefont{Koga}}, \bibinfo {author}
  {\bibfnamefont{N.}~\bibnamefont{Kawakami}}, \bibinfo {author}
  {\bibfnamefont{R.}~\bibnamefont{Peters}},\ and\ \bibinfo {author}
  {\bibfnamefont{T.}~\bibnamefont{Pruschke}},\ }%
  \bibfield{journal}{%
  \bibinfo {journal} {Phys. Rev. B}\ }%
  \textbf{\bibinfo {volume} {77}},\ \bibinfo {pages} {045120} (\bibinfo {year}
  {2008})%
  \bibAnnoteFile{NoStop}{Koga:para}%
\bibitem{Koga:magnetic}%
  \BibitemOpen
  \bibfield{author}{%
  \bibinfo {author} {\bibfnamefont{A.}~\bibnamefont{Koga}}, \bibinfo {author}
  {\bibfnamefont{N.}~\bibnamefont{Kawakami}}, \bibinfo {author}
  {\bibfnamefont{R.}~\bibnamefont{Peters}},\ and\ \bibinfo {author}
  {\bibfnamefont{T.}~\bibnamefont{Pruschke}},\ }%
  \bibfield{journal}{%
  \bibinfo {journal} {J. Phys. Soc. Jpn.}\ }%
  \textbf{\bibinfo {volume} {77}},\ \bibinfo {pages} {033704} (\bibinfo {year}
  {2008})%
  \bibAnnoteFile{NoStop}{Koga:magnetic}%
\bibitem{Zawa:RG}%
  \BibitemOpen
  \bibfield{author}{%
  \bibinfo {author} {\bibfnamefont{C.~M.}\ \bibnamefont{Varma}}\ and\ \bibinfo
  {author} {\bibfnamefont{A.}~\bibnamefont{Zawadowski}},\ }%
  \bibfield{journal}{%
  \bibinfo {journal} {Phys. Rev. B}\ }%
  \textbf{\bibinfo {volume} {32}},\ \bibinfo {pages} {7399} (\bibinfo {year}
  {1985})%
  \bibAnnoteFile{NoStop}{Zawa:RG}%
\bibitem{Solyom:RG}%
  \BibitemOpen
  \bibfield{author}{%
  \bibinfo {author} {\bibfnamefont{K.}~\bibnamefont{Penc}}\ and\ \bibinfo
  {author} {\bibfnamefont{J.}~\bibnamefont{S\'olyom}},\ }%
  \bibfield{journal}{%
  \bibinfo {journal} {Phys. Rev. B}\ }%
  \textbf{\bibinfo {volume} {41}},\ \bibinfo {pages} {704} (\bibinfo {year}
  {1990})%
  \bibAnnoteFile{NoStop}{Solyom:RG}%
\bibitem{White:DMRG1}%
  \BibitemOpen
  \bibfield{author}{%
  \bibinfo {author} {\bibfnamefont{S.~R.}\ \bibnamefont{White}},\ }%
  \bibfield{journal}{%
  \bibinfo {journal} {Phys. Rev. Lett.}\ }%
  \textbf{\bibinfo {volume} {69}},\ \bibinfo {pages} {2863} (\bibinfo {year}
  {1992})%
  \bibAnnoteFile{NoStop}{White:DMRG1}%
\bibitem{White:DMRG2}%
  \BibitemOpen
  \bibfield{author}{%
  \bibinfo {author} {\bibfnamefont{S.~R.}\ \bibnamefont{White}},\ }%
  \bibfield{journal}{%
  \bibinfo {journal} {Phys. Rev. B}\ }%
  \textbf{\bibinfo {volume} {48}},\ \bibinfo {pages} {10345} (\bibinfo {year}
  {1993})%
  \bibAnnoteFile{NoStop}{White:DMRG2}%
\bibitem{schollwock2005}%
  \BibitemOpen
  \bibfield{author}{%
  \bibinfo {author} {\bibfnamefont{U.}~\bibnamefont{Schollw\"ock}},\ }%
  \bibfield{journal}{%
  \bibinfo {journal} {Rev. Mod. Phys.}\ }%
  \textbf{\bibinfo {volume} {77}},\ \bibinfo {pages} {259} (\bibinfo {year}
  {2005})%
  \bibAnnoteFile{NoStop}{schollwock2005}%
\bibitem{manmana2005}%
  \BibitemOpen
  \bibfield{author}{%
  \bibinfo {author} {\bibfnamefont{R.~M.}\ \bibnamefont{Noack}}\ and\ \bibinfo
  {author} {\bibfnamefont{S.~R.}\ \bibnamefont{Manmana}},\ }%
  \bibfield{journal}{%
  \bibinfo {journal} {AIP Conf. Proc.}\ }%
  \textbf{\bibinfo {volume} {789}},\ \bibinfo {pages} {93} (\bibinfo {year}
  {2005})%
  \bibAnnoteFile{NoStop}{manmana2005}%
\bibitem{hallberg2006}%
  \BibitemOpen
  \bibfield{author}{%
  \bibinfo {author} {\bibfnamefont{K.}~\bibnamefont{Hallberg}},\ }%
  \bibfield{journal}{%
  \bibinfo {journal} {Adv. Phys.}\ }%
  \textbf{\bibinfo {volume} {55}},\ \bibinfo {pages} {477} (\bibinfo {year}
  {2006})%
  \bibAnnoteFile{NoStop}{hallberg2006}%
\bibitem{legeza2003b}%
  \BibitemOpen
  \bibfield{author}{%
  \bibinfo {author} {\bibfnamefont{{\"O}.}~\bibnamefont{Legeza}}\ and\ \bibinfo
  {author} {\bibfnamefont{J.}~\bibnamefont{S\'olyom}},\ }%
  \bibfield{journal}{%
  \bibinfo {journal} {Phys. Rev. B}\ }%
  \textbf{\bibinfo {volume} {68}},\ \bibinfo {pages} {195116} (\bibinfo {year}
  {2003})%
  \bibAnnoteFile{NoStop}{legeza2003b}%
\bibitem{vidallatorre03}%
  \BibitemOpen
  \bibfield{author}{%
  \bibinfo {author} {\bibfnamefont{G.}~\bibnamefont{Vidal}}, \bibinfo {author}
  {\bibfnamefont{J.~I.}\ \bibnamefont{Latorre}}, \bibinfo {author}
  {\bibfnamefont{E.}~\bibnamefont{Rico}},\ and\ \bibinfo {author}
  {\bibfnamefont{A.}~\bibnamefont{Kitaev}},\ }%
  \bibfield{journal}{%
  \bibinfo {journal} {Phys. Rev. Lett.}\ }%
  \textbf{\bibinfo {volume} {90}},\ \bibinfo {pages} {227902} (\bibinfo {year}
  {2003})%
  \bibAnnoteFile{NoStop}{vidallatorre03}%
\bibitem{calabrese04}%
  \BibitemOpen
  \bibfield{author}{%
  \bibinfo {author} {\bibfnamefont{P.}~\bibnamefont{Calabrese}}\ and\ \bibinfo
  {author} {\bibfnamefont{J.}~\bibnamefont{Cardy}},\ }%
  \bibfield{journal}{%
  \bibinfo {journal} {J. Stat. Mech.}\ }%
  \textbf{\bibinfo {volume} {2004}},\ \bibinfo {pages} {P06002} (\bibinfo
  {year} {2004})%
  \bibAnnoteFile{NoStop}{calabrese04}%
\bibitem{rissler2006}%
  \BibitemOpen
  \bibfield{author}{%
  \bibinfo {author} {\bibfnamefont{J.}~\bibnamefont{Rissler}}, \bibinfo
  {author} {\bibfnamefont{R.~M.}\ \bibnamefont{Noack}},\ and\ \bibinfo {author}
  {\bibfnamefont{S.~R.}\ \bibnamefont{White}},\ }%
  \bibfield{journal}{%
  \bibinfo {journal} {Chem. Phys.}\ }%
  \textbf{\bibinfo {volume} {323}},\ \bibinfo {pages} {519 } (\bibinfo {year}
  {2006})%
  \bibAnnoteFile{NoStop}{rissler2006}%
\bibitem{legeza2006}%
  \BibitemOpen
  \bibfield{author}{%
  \bibinfo {author} {\bibfnamefont{{\"O}.}~\bibnamefont{Legeza}}\ and\ \bibinfo
  {author} {\bibfnamefont{J.}~\bibnamefont{S\'olyom}},\ }%
  \bibfield{journal}{%
  \bibinfo {journal} {Phys. Rev. Lett.}\ }%
  \textbf{\bibinfo {volume} {96}},\ \bibinfo {pages} {116401} (\bibinfo {year}
  {2006})%
  \bibAnnoteFile{NoStop}{legeza2006}%
\bibitem{luigi2008}%
  \BibitemOpen
  \bibfield{author}{%
  \bibinfo {author} {\bibfnamefont{L.}~\bibnamefont{Amico}}, \bibinfo {author}
  {\bibfnamefont{R.}~\bibnamefont{Fazio}}, \bibinfo {author}
  {\bibfnamefont{A.}~\bibnamefont{Osterloh}},\ and\ \bibinfo {author}
  {\bibfnamefont{V.}~\bibnamefont{Vedral}},\ }%
  \bibfield{journal}{%
  \bibinfo {journal} {Rev. Mod. Phys.}\ }%
  \textbf{\bibinfo {volume} {80}},\ \bibinfo {pages} {517} (\bibinfo {year}
  {2008})%
  \bibAnnoteFile{NoStop}{luigi2008}%
\bibitem{gu:prl2004}%
  \BibitemOpen
  \bibfield{author}{%
  \bibinfo {author} {\bibfnamefont{S.-J.}\ \bibnamefont{Gu}}, \bibinfo {author}
  {\bibfnamefont{S.-S.}\ \bibnamefont{Deng}}, \bibinfo {author}
  {\bibfnamefont{Y.-Q.}\ \bibnamefont{Li}},\ and\ \bibinfo {author}
  {\bibfnamefont{H.-Q.}\ \bibnamefont{Lin}},\ }%
  \bibfield{journal}{%
  \bibinfo {journal} {Phys. Rev. Lett.}\ }%
  \textbf{\bibinfo {volume} {93}},\ \bibinfo {pages} {086402} (\bibinfo {year}
  {2004})%
  \bibAnnoteFile{NoStop}{gu:prl2004}%
\bibitem{wu:prl2004}%
  \BibitemOpen
  \bibfield{author}{%
  \bibinfo {author} {\bibfnamefont{L.-A.}\ \bibnamefont{Wu}}, \bibinfo {author}
  {\bibfnamefont{M.~S.}\ \bibnamefont{Sarandy}},\ and\ \bibinfo {author}
  {\bibfnamefont{D.~A.}\ \bibnamefont{Lidar}},\ }%
  \bibfield{journal}{%
  \bibinfo {journal} {Phys. Rev. Lett.}\ }%
  \textbf{\bibinfo {volume} {93}},\ \bibinfo {pages} {250404} (\bibinfo {year}
  {2004})%
  \bibAnnoteFile{NoStop}{wu:prl2004}%
\bibitem{yang:pra2005}%
  \BibitemOpen
  \bibfield{author}{%
  \bibinfo {author} {\bibfnamefont{M.-F.}\ \bibnamefont{Yang}},\ }%
  \bibfield{journal}{%
  \bibinfo {journal} {Phys. Rev. A}\ }%
  \textbf{\bibinfo {volume} {71}},\ \bibinfo {pages} {030302} (\bibinfo {year}
  {2005})%
  \bibAnnoteFile{NoStop}{yang:pra2005}%
\bibitem{deng:prb2006}%
  \BibitemOpen
  \bibfield{author}{%
  \bibinfo {author} {\bibfnamefont{S.-S.}\ \bibnamefont{Deng}}, \bibinfo
  {author} {\bibfnamefont{S.-J.}\ \bibnamefont{Gu}},\ and\ \bibinfo {author}
  {\bibfnamefont{H.-Q.}\ \bibnamefont{Lin}},\ }%
  \bibfield{journal}{%
  \bibinfo {journal} {Phys. Rev. B}\ }%
  \textbf{\bibinfo {volume} {74}},\ \bibinfo {pages} {045103} (\bibinfo {month}
  {Jul}\ \bibinfo {year} {2006})%
  \bibAnnoteFile{NoStop}{deng:prb2006}%
\bibitem{DBSS:cikk1}%
  \BibitemOpen
  \bibfield{author}{%
  \bibinfo {author} {\bibfnamefont{{\"O}.}~\bibnamefont{Legeza}}, \bibinfo
  {author} {\bibfnamefont{J.}~\bibnamefont{R\"oder}},\ and\ \bibinfo {author}
  {\bibfnamefont{B.~A.}\ \bibnamefont{Hess}},\ }%
  \bibfield{journal}{%
  \bibinfo {journal} {Phys. Rev. B}\ }%
  \textbf{\bibinfo {volume} {67}},\ \bibinfo {pages} {125114} (\bibinfo {year}
  {2003})%
  \bibAnnoteFile{NoStop}{DBSS:cikk1}%
\bibitem{DBSS:cikk2}%
  \BibitemOpen
  \bibfield{author}{%
  \bibinfo {author} {\bibfnamefont{{\"O}.}~\bibnamefont{Legeza}}\ and\ \bibinfo
  {author} {\bibfnamefont{J.}~\bibnamefont{S\'olyom}},\ }%
  \bibfield{journal}{%
  \bibinfo {journal} {Phys. Rev. B}\ }%
  \textbf{\bibinfo {volume} {70}},\ \bibinfo {pages} {205118} (\bibinfo {year}
  {2004})%
  \bibAnnoteFile{NoStop}{DBSS:cikk2}%
\bibitem{barcza2010}%
  \BibitemOpen
  \bibfield{author}{%
  \bibinfo {author} {\bibfnamefont{G.}~\bibnamefont{Barcza}}, \bibinfo {author}
  {\bibfnamefont{{\"O}.}~\bibnamefont{Legeza}}, \bibinfo {author}
  {\bibfnamefont{K.~H.}\ \bibnamefont{Marti}},\ and\ \bibinfo {author}
  {\bibfnamefont{M.}~\bibnamefont{Reiher}},\ }%
  \bibfield{journal}{%
  \bibinfo {journal} {Phys. Rev. A}\ }%
  \textbf{\bibinfo {volume} {83}},\ \bibinfo {pages} {012508} (\bibinfo {year}
  {2011})%
  \bibAnnoteFile{NoStop}{barcza2010}%
\bibitem{Boguslawski}%
  \BibitemOpen
  \bibfield{author}{%
  \bibinfo {author} {\bibfnamefont{K.}~\bibnamefont{Boguslawski}}, \bibinfo
  {author} {\bibfnamefont{P.}~\bibnamefont{Tecmer}}, \bibinfo {author}
  {\bibfnamefont{G.}~\bibnamefont{Barcza}}, \bibinfo {author}
  {\bibfnamefont{{\"O}.}~\bibnamefont{Legeza}},\ and\ \bibinfo {author}
  {\bibfnamefont{M.}~\bibnamefont{Reiher}},\ }%
  \bibfield{journal}{%
  \bibinfo {journal} {J. Chem. Theory Comp.}\ }%
  \textbf{\bibinfo {volume} {9}},\ \bibinfo {pages} {2959} (\bibinfo {year}
  {2013})%
  \bibAnnoteFile{NoStop}{Boguslawski}%
\bibitem{wolf2008}%
  \BibitemOpen
  \bibfield{author}{%
  \bibinfo {author} {\bibfnamefont{M.~M.}\ \bibnamefont{Wolf}}, \bibinfo
  {author} {\bibfnamefont{F.}~\bibnamefont{Verstraete}}, \bibinfo {author}
  {\bibfnamefont{M.~B.}\ \bibnamefont{Hastings}},\ and\ \bibinfo {author}
  {\bibfnamefont{J.~I.}\ \bibnamefont{Cirac}},\ }%
  \bibfield{journal}{%
  \bibinfo {journal} {Phys. Rev. Lett.}\ }%
  \textbf{\bibinfo {volume} {100}},\ \bibinfo {pages} {070502} (\bibinfo {year}
  {2008})%
  \bibAnnoteFile{NoStop}{wolf2008}%
\bibitem{furukawa2009}%
  \BibitemOpen
  \bibfield{author}{%
  \bibinfo {author} {\bibfnamefont{S.}~\bibnamefont{Furukawa}}, \bibinfo
  {author} {\bibfnamefont{V.}~\bibnamefont{Pasquier}},\ and\ \bibinfo {author}
  {\bibfnamefont{J.}~\bibnamefont{Shiraishi}},\ }%
  \bibfield{journal}{%
  \bibinfo {journal} {Phys. Rev. Lett.}\ }%
  \textbf{\bibinfo {volume} {102}},\ \bibinfo {pages} {170602} (\bibinfo {year}
  {2009})%
  \bibAnnoteFile{NoStop}{furukawa2009}%
\bibitem{legeza:entanglement}%
  \BibitemOpen
  \bibfield{author}{%
  \bibinfo {author} {\bibfnamefont{G.}~\bibnamefont{Barcza}}, \bibinfo {author}
  {\bibfnamefont{R.~M.}\ \bibnamefont{Noack}}, \bibinfo {author}
  {\bibfnamefont{J.}~\bibnamefont{S\'olyom}},\ and\ \bibinfo {author}
  {\bibfnamefont{{\"O}.}~\bibnamefont{Legeza}},\ }%
  \bibinfo {journal} {arXiv:1406.6643}%
  \bibAnnoteFile{NoStop}{legeza:entanglement}%
\bibitem{White:S1chain}%
  \BibitemOpen
\bibfield{journal}{%
    }%
  \bibfield{author}{%
  \bibinfo {author} {\bibfnamefont{S.~R.}\ \bibnamefont{White}}\ and\ \bibinfo
  {author} {\bibfnamefont{D.~A.}\ \bibnamefont{Huse}},\ }%
  \bibfield{journal}{%
  \bibinfo {journal} {Phys. Rev. B}\ }%
  \textbf{\bibinfo {volume} {48}},\ \bibinfo {pages} {3844} (\bibinfo {year}
  {1993})%
  \bibAnnoteFile{NoStop}{White:S1chain}%
\bibitem{legeza:ladder}%
  \BibitemOpen
  \bibfield{author}{%
  \bibinfo {author} {\bibfnamefont{E.~H.}\ \bibnamefont{Kim}}, \bibinfo
  {author} {\bibfnamefont{{\"O}.}~\bibnamefont{Legeza}},\ and\ \bibinfo
  {author} {\bibfnamefont{J.}~\bibnamefont{S\'olyom}},\ }%
  \bibfield{journal}{%
  \bibinfo {journal} {Phys. Rev. B}\ }%
  \textbf{\bibinfo {volume} {77}},\ \bibinfo {pages} {205121} (\bibinfo {year}
  {2008})%
  \bibAnnoteFile{NoStop}{legeza:ladder}%
\bibitem{Haldane:entanglement}%
  \BibitemOpen
  \bibfield{author}{%
  \bibinfo {author} {\bibfnamefont{H.}~\bibnamefont{Li}}\ and\ \bibinfo
  {author} {\bibfnamefont{F.~D.~M.}\ \bibnamefont{Haldane}},\ }%
  \bibfield{journal}{%
  \bibinfo {journal} {Phys. Rev. Lett.}\ }%
  \textbf{\bibinfo {volume} {101}},\ \bibinfo {pages} {010504} (\bibinfo {year}
  {2008})%
  \bibAnnoteFile{NoStop}{Haldane:entanglement}%
\bibitem{Pollmann:entanglement}%
  \BibitemOpen
  \bibfield{author}{%
  \bibinfo {author} {\bibfnamefont{F.}~\bibnamefont{Pollmann}}, \bibinfo
  {author} {\bibfnamefont{A.~M.}\ \bibnamefont{Turner}}, \bibinfo {author}
  {\bibfnamefont{E.}~\bibnamefont{Berg}},\ and\ \bibinfo {author}
  {\bibfnamefont{M.}~\bibnamefont{Oshikawa}},\ }%
  \bibfield{journal}{%
  \bibinfo {journal} {Phys. Rev. B}\ }%
  \textbf{\bibinfo {volume} {81}},\ \bibinfo {pages} {064439} (\bibinfo {year}
  {2010})%
  \bibAnnoteFile{NoStop}{Pollmann:entanglement}%
\bibitem{Sirker:dimerization}%
  \BibitemOpen
  \bibfield{author}{%
  \bibinfo {author} {\bibfnamefont{J.}~\bibnamefont{Sirker}}, \bibinfo {author}
  {\bibfnamefont{A.}~\bibnamefont{Herzog}}, \bibinfo {author}
  {\bibfnamefont{A.~M.}\ \bibnamefont{Ole\ifmmode~\acute{s}\else \'{s}\fi{}}},\
  and\ \bibinfo {author} {\bibfnamefont{P.}~\bibnamefont{Horsch}},\ }%
  \bibfield{journal}{%
  \bibinfo {journal} {Phys. Rev. Lett.}\ }%
  \textbf{\bibinfo {volume} {101}},\ \bibinfo {pages} {157204} (\bibinfo {year}
  {2008})%
  \bibAnnoteFile{NoStop}{Sirker:dimerization}%
\end{thebibliography}%

\end{document}